\documentclass[12pt]{article}

\usepackage {epsfig}

\usepackage{color}
\usepackage{graphicx}

\newcommand{\ba}{\begin{eqnarray}}
\newcommand{\ea}{\end{eqnarray}}
\newcommand{\be}{\begin{equation}}
\newcommand{\ee}{\end{equation}}

\begin{document}

\title{Two-Photon Physics in Hadronic Processes}

%\vspace*{0.5cm}

\author{Carl E. Carlson\\
Department of Physics\\
College of William and Mary,
Williamsburg, Virginia 23187\\
~~\\
and\\
~~\\
Marc Vanderhaeghen\\
Thomas Jefferson National Accelerator Facility\\
Newport News, VA 23606\\
and\\
Department of Physics\\
College of William and Mary, Williamsburg, Virginia 23187}

\date{31 January 2007}

\maketitle

{\sc key words}:\quad  %list 3 to 5 key words that are not in title

electron scattering, form factors, two-photon exchange processes

%\bigskip
\medskip
\hrule
%\bigskip
\begin{abstract}

Two-photon exchange contributions to elastic electron-scattering
are reviewed. The apparent discrepancy in the extraction of elastic
nucleon form factors between unpolarized Rosenbluth and
polarization transfer experiments is discussed, as well as
the understanding of
this puzzle in terms of two-photon exchange corrections.
Calculations of such corrections both within partonic and hadronic
frameworks are reviewed.
In view of recent spin-dependent electron scattering data,
the relation of the two-photon exchange process
to the hyperfine splitting in hydrogen is critically examined.
The imaginary part of the two-photon exchange amplitude as can be
accessed from the beam normal spin asymmetry in elastic electron-nucleon
scattering is reviewed. Further extensions and open issues in this field 
are outlined. 

\end{abstract}

%\bigskip
\hrule
%\bigskip

\begin{quote}
\hskip 24 pt
I like a thing simple but it must be simple through complication.  Everything must come into your scheme, otherwise you cannot achieve real simplicity.

\hfill Gertrude Stein, 
``Afterword,'' {\it What Are Masterpieces}
\end{quote}

%\bigskip
\newpage
{\small \tableofcontents }

\bigskip

%%%%%%%%%%%%%%%%%%%%%%%%%
% baselineskip chosen to give about 430 words per page (CEC, Jan 2006)
\baselineskip 13.3pt
%%%%%%%%%%%%%%%%%%%%%%%%%

%%%%%%%%%%%%%%%%%%%%%%%%%%%%%%%%%%%%%%%%%%%%%%%%%%%%%%%%%%%%%%%%%%%%%%%%%%

\section{Introduction}
\label{sec1}

%%%%%%%%%%%%%%%%%%%%%%%%%%%%%%%%%%%%%%%%%%%%%%%%%%%%%%%%%%%%%%%%%%%%%%%%%%

Elastic electron-nucleon scattering in the one-photon exchange approximation 
is a time-honored tool to access information on the structure of hadrons. 
Experiments with increasing precision have become possible in recent years, 
mainly triggered by new techniques to perform polarization experiments 
at electron scattering facilities. This opened a new 
frontier in the measurement of hadron structure quantities, such as 
its electroweak form factors, parity violating effects, 
nucleon polarizabilities, transition form factors, or 
the measurement of spin dependent structure functions, to name a few. 
For example, experiments using 
polarized electron beams and measuring the ratio of the recoil nucleon 
in-plane polarization components have profoundly extended our understanding 
of the nucleon electromagnetic form factors (FFs);  
for recent reviews on nucleon FFs see e.g. 
Refs.~\cite{Hyde-Wright:2004gh,Arrington:2006zm,Perdrisat:2006hj}. 
For the proton, such polarization experiments access 
the ratio $G_{E} / G_{M}$ of the proton's electric ( $G_{E}$ ) to 
magnetic ( $G_{M}$ ) FFs directly from the ratio 
of the ``sideways'' and ``longitudinal'' polarizations 
in elastic electron-nucleon scattering as~\cite{Arn81}, 
\ba
\label{eq:polratio}
{P_s \over P_l} = 
		- \sqrt{2\varepsilon \over \tau (1+\varepsilon)}\  
		{ G_E(Q^2)  \over G_M(Q^2)
		}  \ .
\ea
Here,
\ba
 \tau \equiv {Q^2 \over 4M^2}   \ , \qquad 
{1 \over \varepsilon} \equiv 1 + 2(1+\tau) \tan^2 {\theta\over 2} \ ,
\ea
$Q^2=-q^2$ is the momentum transfer squared, $\theta$ is the laboratory
scattering angle, and $0 \le \varepsilon \le 1$. 
Recently, this ratio has been measured at the Jefferson Laboratory 
out to a space-like momentum transfer $Q^2$ of 
5.6 GeV$^2$~\cite{Jones00,Gayou02,Punjabi:2005wq} . 
It came as a surprise that these experiments 
extracted a ratio of $G_{E} / G_{M}$ which is clearly at variance with 
unpolarized measurements \cite{Slac94,Chr04,Qattan:2004ht} 
using the Rosenbluth separation technique, which 
measures the angular dependence of the differential 
cross section for elastic electron-nucleon scattering at fixed $Q^2$,
\ba
\label{eq:rosie}
{d\sigma \over d\Omega_{Lab}} \propto
		 G_M^2 + \frac{\varepsilon}{\tau} G_E^2 
				\ ,
\ea
where the proportionality factor is well known, 
and isolates the $\varepsilon$-dependent term. 
In each case, the quoted formulas assume single-photon exchange 
between electron and nucleon.

To explain the discrepancy between the two experimental techniques, 
suspicion falls on the Rosenbluth measurements, but not because 
of experimental problems $per se$.  
The Rosenbluth formula, Eq.~(\ref{eq:rosie}), 
at high $Q^2$ has a numerically big term, $G_M^2$, and a small term.  
The results for $G_E$ come from the small term.  Any omitted
$\varepsilon$-dependent corrections to the large term can thus have a
strikingly large effect on $G_E$.  

\begin{figure}[b]%[htbp]
\begin{center}

\centerline{ \includegraphics[height=23mm]{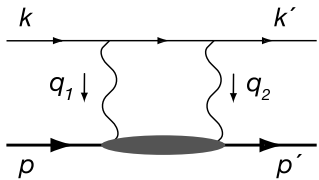} }

\caption{Two photon exchange in the lepton - nucleon  
scattering process $l (k) + N (p) \to l (k^\prime) + N(p^\prime)$, 
with $k, k^\prime$ ($p, p^\prime$) the four-momenta of leptons (nucleons) 
respectively.}
\label{twophotonscatt}
\end{center}
\end{figure}

Two-photon exchange, Fig.~\ref{twophotonscatt}, is one thinkable culprit.  
The subject has a long history. In fact as early as the late 1950s 
and during the 1960s, when 
electron-nucleon scattering was measured systematically at the Stanford Linear
Accelerator Center to access nucleon electromagnetic FFs, the
validity of the one-photon exchange approximation has been discussed both 
theoretically and experimentally. Because the one-photon exchange cross
section depends quadratically on the lepton charge, the difference between
electron-nucleon and positron-nucleon cross sections is a test for two- or
multi-photon exchange processes. Early comparisons of electron- and
positron-nucleon scattering cross sections were consistent with equal cross
sections~\cite{Mar68,Hartwig:1975px}, 
but the precision achieved in those early investigations could not exclude 
two-photon exchange effects at the few percent level of the cross section. 
Contributions of this size can be expected due to the additional
electromagnetic coupling in the two-photon exchange diagram, 
which brings in a suppression factor $\alpha = e^2 / (4 \pi) \simeq 1/ 137$.

%%%%%%%%%%%%%%%%%%%%%%%%%%%%%%%

%%%%%%%%%%%%%%%%%%%%%%%%%%%%%%%

On the theory side, the calculation of corrections to elastic electron-nucleon 
scattering of order $e^2$ relative to the Born approximation, also known 
as radiative corrections, have a long history, see 
{\it e.g.} Refs.~\cite{MoTsai68,oldyennie}, which were heavily applied in the
analysis of early electron-nucleon scattering experiments.    
The infrared divergences associated with one photon 
in Fig.~\ref{twophotonscatt} being soft, 
{\it i.e.}, having a vanishingly small four-momentum, 
had been extracted---they cancel against infrared divergences 
from soft bremsstrahlung---but the contributions when both 
photons are hard ({\it i.e.}, when the momentum transfer of both 
individual photons is large) were not calculated in those works 
because of insufficient knowledge of the intermediate hadron state. 
The early calculators were well aware of the omission, 
and explicitly expressed hope that the missing contributions would be small.
Some early estimates of the two-photon exchange  
contribution of Fig.~\ref{twophotonscatt} with two hard photons 
were attempted though in the late 1950s by Drell and 
collaborators~\cite{drell57,drell59}. 
Those works constructed a non-relativistic model 
for the blob in Fig.~\ref{twophotonscatt}, 
including the nucleon and lowest nucleon resonance contribution, 
the $\Delta(1232)$. The calculation found that the resonance contribution to
the two-photon exchange diagram affects the cross sections at the 
$\sim 1$~\% level. Due to the non-relativistic nature of the calculation, 
the result was limited to about 1 GeV electron beam energy. 
In subsequent works, e.g. Refs.~\cite{werthamer61, Greenhut:1970aq},
two-photon exchange effects were approximately calculated to higher
energies. In particular, Greenhut~\cite{Greenhut:1970aq} evaluated 
the contribution of higher nucleon resonance intermediate states, with masses 
up to 1.7 GeV, when evaluating the blob in Fig.~\ref{twophotonscatt}.  
It was found that the dispersive (real) part 
of the two-photon amplitude yields an electron to positron cross section ratio
which deviates from unity at the 1-2 \% level in the few GeV region. 
The relative smallness of the resonance contribution partly originates because
the real parts of the resonance amplitudes change sign in the integrand
entering the evaluation of the box diagram. 
 
Triggered by the experimental discrepancy 
between polarization transfer and Rosenbluth
measurements of the proton form factor ratio $G_{E}/G_{M}$ at larger 
$Q^2$ in recent years, the field has seen a new life. 
In 2003, it was noticed in Ref.~\cite{GV03} 
that the general form of the two-photon
exchange graphs could be expressed in an effective current $\times$ current
form, but with an extra structure beyond those that gave $G_{M }$ 
and $G_{E }$.
Further, if this extra term had the size one might estimate from perturbation
theory, then its interference with the one-photon exchange amplitude 
could be comparable in size to the $(G_{E })^2$ term in the Rosenbluth cross
section at large $Q^2$.  In addition, there could be $\varepsilon$-dependent 
modifications to the $(G_{M })^2$ term.  
Hence there was motivation for a precise evaluation.
Realistic calculations of
elastic electron-nucleon scattering beyond the Born approximation are
required in order to demonstrate in a quantitative way that 
$2 \gamma$ exchange effects are indeed able to resolve this 
discrepancy at larger $Q^2$. 
In the current work, we will review several such attempts 
and describe the present status of this field. 
On the experimental side, recent years have seen first attempts to 
extract the effect of two-photon exchange contributions 
in a quantitative way from the electron-proton scattering
data~\cite{Arrington}. 

Besides offering a way to explain the glaring discrepancy between 
two methods of measuring the proton electric form factor, 
the study of two-photon processes also contain opportunities to 
access nucleon structure physics which surpasses the information contained in
nucleon FFs. The possibility arises  
because a successful two-photon calculation involving a hadronic 
system requires knowledge of hadronic structure, 
of a sort which has only been available recently.  
For example, one line of investigation arises when the virtuality 
of one or both of the photons in the two-photon process 
is large compared to a nucleon mass scale. In that case, 
the hard scale allows one to access the Compton scattering 
subprocess on a quark within the nucleon. 
The new (non-perturbative) pieces of information which one 
then accesses from such a process are the quark correlation 
functions within the nucleon, also known as 
generalized parton distributions (GPDs); for recent reviews see
Refs.~\cite{Ji:1998pc,GPV01,Diehl:2003ny,Ji:2004gf,Belitsky:2005qn}.  
We will review how these GPDs can in turn be used to estimate the 
two-photon exchange diagram of Fig.~\ref{twophotonscatt} at large $Q^2$. 

Another line of calculations involves the hadronic corrections to 
ultra-high precision atomic physics experiments, such as the 
hydrogen hyperfine-splitting. Theoretical predictions for quantities 
such as the Lamb shift in hydrogen or the hydrogen hyperfine splitting 
can currently be performed within quantum electrodynamics to 
such accuracy that the leading theoretical uncertainties are related to 
the nuclear size or the nuclear excitation spectrum;   
for reviews see e.g. Refs.~\cite{Eides:2000xc,Karshenboim:2005iy}. 
For the hyperfine splitting in hydrogen, which is known to 13
significant figures, current theoretical understanding is 
at the part per million level. The leading theoretical uncertainty involves  
the calculation of the two-photon exchange graph 
of Fig.~\ref{twophotonscatt} for zero momentum transfer 
between the bound electron and the proton, 
allowing for all nucleon excited states in the blob. The
current theoretical understanding of these two-photon 
exchange corrections will be reviewed in this work.  

There are several physical problems where a one-photon exchange potential 
is not sufficiently accurate. Besides the description of simple atoms, such as
hydrogen and helium, also for a precise description of positronium one needs
to include two- and multi-photon exchange effects. In particular in the 
interaction of electrically neutral 
systems, such as neutral atoms and molecules, the effect of two-photon
exchange gives the dominant contribution to the forces between such systems,
for a theoretical review of such dispersion forces; see
{\it e.g.} Ref.~\cite{Feinberg:1989ps}.  

To push the precision frontier further in electron scattering as well as in
the hadronic corrections to atomic physics quantities, one needs 
a good control of $2 \gamma$ exchange mechanisms 
and needs to understand how they may or may not affect 
different observables. This justifies a systematic study of such 
$2 \gamma$ exchange effects, both theoretically and experimentally.  
Besides the real (dispersive) part of the $2 \gamma$ exchange amplitude, 
which can be accessed by reversing the sign of the lepton charge, 
also precise measurements of the imaginary part 
of the $2 \gamma$ exchange amplitude became possible in very recent years. 
The imaginary (absorptive) part of the $2 \gamma$ exchange amplitude 
can be accessed through a single spin asymmetry (SSA) in 
elastic electron-nucleon scattering, when either the target or beam spin 
are polarized normal to the scattering plane, as has been discussed some time 
ago~\cite{Barut60,RKR71}. As time reversal invariance forces 
this SSA to vanish for one-photon exchange, it 
is of order $\alpha = e^2 / (4 \pi) \simeq 1/ 137$. Furthermore, to polarize   
an ultra-relativistic particle in the direction 
normal to its momentum involves 
a suppression factor $m / E$ (with $m$ the mass and $E$ the 
energy of the particle), 
which typically is of order $10^{-4} - 10^{-3}$ 
for the electron when the electron 
beam energy is in the 1 GeV range. Therefore, the 
resulting target normal SSA can be expected to be of order $10^{-2}$, 
whereas the beam normal SSA is of order $10^{-6} - 10^{-5}$.
Measurements of small asymmetries of the order ppm 
are quite demanding experimentally, but have been performed in very recent 
years, and will also be reviewed in this work. 

The outline of the present work is as follows. In Section~2, we review the 
elastic electron-nucleon scattering beyond the Born approximation and
highlight the discrepancy in the extraction of $G_E / G_M$ using polarization
transfer and unpolarized (Rosenbluth) measurements. We give a brief review of
the different attempts which have been made recently 
to explain this difference in
terms of two-photon exchange corrections, and present in more detail 
a partonic description at larger momentum transfers. 

In Section~3, we describe the hadronic corrections of the hydrogen hyperfine
splitting, based on the latest evaluation of the forward polarized structure
functions which enter in the calculation of the two-photon exchange diagram. 

In Section~4, we review the beam and target single spin asymmetries which
measure the imaginary part of the two-photon exchange amplitude. In particular,
we give an overview of the recent high precision measurements in case of a
polarized beam with normal beam spin polarization. 

We conclude in Section~5, and spell out a few open issues in this
field.
%%%%%%%%%%%%%%%%%%%%%%%%%%%%%%%%%%%%%%%%%%%%%%%%%%%%%%%%%%%%%%%%%%%%%%%%%%

\section{Two-photon in elastic electron-nucleon scattering}

%%%%%%%%%%%%%%%%%%%%%%%%%%%%%%%%%%%%%%%%%%%%%%%%%%%%%%%%%%%%%%%%%%%%%%%%%%

%%%%%%%%%%%%%%%%%%%%%%%%%%%%%%%%%%%%%%%%%%%%%%%%%%%%%%%%%%%%%%%%%%%%%%%%%%

\subsection{Hard two-photon exchange}

%%%%%%%%%%%%%%%%%%%%%%%%%%%%%%%%%%%%%%%%%%%%%%%%%%%%%%%%%%%%%%%%%%%%%%%%%%

In this section, we study attempts to evaluate completely the two-photon exchange contributions to electron-proton elastic scattering, specifically including the exchange of two hard photons, which can probe well inside the proton and which require detailed knowledge of proton structure to evaluate.  The immediate motivation, already given in the introduction, is the conflict between the Rosenbluth, with pre-2003 radiative corrections, and the polarization measurements of $G_E/G_M$ for the proton.  The difference between the two techniques is a factor 4 at $Q^2 = 5.6$ GeV$^2$!  (The data will appear on plots later in this section,  when we discuss how well the proposed resolutions of the conundrum are working.)

Modern quantitative calculations either treat the hadronic intermediate state in Fig.~\ref{twophotonscatt} as a proton plus a set of resonances, or else treat it in a constituent picture using generalized parton distributions.

In the earliest of the modern calculations, Blunden, Melnitchouk, and Tjon~\cite{BMT03} evaluated the two-photon exchange amplitude keeping just the elastic nucleon intermediate state. They found that the two-photon exchange correction with an intermediate nucleon has the proper sign and magnitude to partially resolve the discrepancy between the two experimental techniques.  Later, the same group, joined by Kondratyuk~\cite{Kondratyuk:2005kk}, included contributions of the $\Delta(1232)$ in the intermediate state, which partly canceled the elastic terms.  Most recently, Kondratyuk and Blunden~\cite{Kondratyuk:2007hc} included five more baryon resonances in the intermediate state.  While finding that the overall contribution of the additional resonances was not large, the totality of their corrections with their choices for the $\gamma$-nucleon-resonance vertices leads to good agreement with the Rosenbluth data using the form factors obtained from polarization data.

Borisyuk and Kobushkin~\cite{Borisyuk:2006fh} also considered two-photon corrections with elastic nucleon intermediate states, and used dispersive techniques to reduce the necessary integrals to ones involving the vertex form factors with only spacelike momentum transfers.  They were able to reduce it further to a single numerical integral for sufficiently low $Q^2$.  (They do not show any Rosenbluth-type plots, but their plotted results for (say) $\delta G_M/G_M$, in notation defined below, are in line with results known from what will be described in the rest of this section, despite the rather different methodology.)

In~\cite{YCC04,ABCCV05}, a group including the present authors calculated the hard two-photon elastic electron-nucleon scattering amplitude at large
momentum transfers by relating the required virtual Compton process on the nucleon to
generalized parton distributions (GPDs), which also enter in other wide
angle scattering processes.  This approach effectively sums all possible excitations of inelastic
nucleon intermediate states.  It was found that the two-photon corrections
to the Rosenbluth process indeed can substantially reconcile the two
ways of measuring $G_E/G_M$.  

Rosenbluth data is also available where the recoiling proton, rather than the electron, is detected~\cite{Qattan:2004ht}.  These data appear to match the data where the scattered electron was detected. The two-photon exchange contributions are the same whatever particle is detected.  However, the bremsstrahlung corrections are different.  We shall defer detailed discussion of the proton-detected data pending reassessment of the original~\cite{oldyennie,oldkrass} and the new ~\cite{Ent:2001hm,Qattan:2004ht} proton-observed bremsstrahlung calculations.

%%%%%%%%%%%%%%%%%%%%%%%%%%%%%%%%%%%%%%%%%%%%%%%%%%%%%%%%%%%%%%%%%%%%%%%%%%

\subsection{Elastic electron-nucleon scattering observables}

\label{sec:observables}

%%%%%%%%%%%%%%%%%%%%%%%%%%%%%%%%%%%%%%%%%%%%%%%%%%%%%%%%%%%%%%%%%%%%%%%%%%

In order to describe elastic electron-nucleon
scattering,
\begin{equation}
										\label{Eq:intro.2}
	l(k,h)+N(p,\lambda_N)\rightarrow l(k',h')+N(p',\lambda'_N),
\end{equation}

\noindent where $h$, $h'$, $\lambda_N$, and $\lambda'_N$ are helicities, we adopt the definitions
\begin{equation}
										\label{Eq:intro.3}
	P=\frac{p+p'}{2},\, K=\frac{k+k'}{2},\, q=k-k'=p'-p \ ,
\end{equation}

\noindent  define the Mandelstam variables
\ba
s = (p+k)^2,  \quad
t = q^2 = - Q^2,   \quad
u = (p-k')^2,
\ea

\noindent let $\nu \equiv K\cdot P$, and let $M$ be the nucleon mass.  

The $T$-matrix helicity amplitudes are given by
\ba
T^{h',h}_{\lambda'_N, \lambda_N} \equiv
	\left\langle k', h'; p', \lambda'_N \right| T 
		\left| k, h; p, \lambda_N \right\rangle  \ .
\ea
Parity invariance reduces the number of independent helicity amplitudes from 16 to 8.  Time reversal invariance further reduces the number to 6~\cite{Goldb57}.  Further still, in a gauge theory lepton helicity is conserved to all orders in perturbation theory when the lepton mass is zero.  We shall neglect the lepton mass.  This finally reduces the number of independent helicity amplitudes to 3, which one may for example choose as
\ba
T^{+,+}_{+,+} \ ; \quad T^{+,+}_{-,-} \ ; \quad 
	T^{+,+}_{-,+} = T^{+,+}_{+,-} \ .
\ea
(The phase in the last equality is for particle momenta in the $xz$ plane.)

Alternatively, one can expand in terms of a set of three independent Lorentz structures, multiplied by three generalized form factors.  One such $T$-matrix expansion is
\ba														\label{eq:tmatrix}
T_{h, \, \lambda'_N \lambda_N} \,&=&\, 
\frac{e^{2}}{Q^{2}} \, \bar{u}(k', h)\gamma _{\mu }u(k, h)\, \\
&\times& \, 
\bar{u}(p', \lambda'_N)\left( \tilde{G}_{M}\, \gamma ^{\mu }
-\tilde{F}_{2}\frac{P^{\mu }}{M}
+\tilde{F}_{3}\frac{\gamma\cdot K P^{\mu }}{M^{2}}\right) u(p, \lambda_N) \ . \nonumber
\ea
The expansion is general.  The overall factors and the notations $\tilde G_M$
and $\tilde F_2$ have been chosen~\cite{GV03} to have a straightforward
connection to the standard form factors in the one-photon exchange limit. 

If desired, one may replace the $\tilde F_3$ term by an axial-like term using the identity,
\ba														\label{eq:theorem}
\bar{u}(k') \gamma\cdot P u(k)  \times  \bar{u}(p') \gamma\cdot K u(p)
		&=&
		{s-u\over 4} \,
		\bar{u}(k') \gamma_\mu u(k)  \times  \bar{u}(p') \gamma^\mu u(p)
								\nonumber \\
		&+&  {t\over 4} \,
		\bar{u}(k') \gamma_\mu \gamma_5 u(k)  \times 
		\bar{u}(p') \gamma^\mu \gamma^5 u(p) \ ,
\ea
which is valid for massless leptons and any nucleon mass.  We will, however,
continue with the $T$-matrix in the form shown in Eq.~(\ref{eq:tmatrix}). An
equivalent expansion has also been studied in Ref.~\cite{Rekalo:2003xa}. 

The scalar quantities $\tilde{G}_{M}$, $\tilde{F}_{2}$, and
$\tilde{F}_{3}$  are complex functions of two variables, say $\nu$ and
$Q^{2}$.  We also use
\ba
\tilde{G}_{E}\equiv\tilde{G}_{M}-(1+\tau )\tilde{F}_{2}  \ .
\ea
To separately identify the one- and two-photon exchange contributions, we
use the notation $\tilde G_M = G_M + \delta \tilde G_M$, and
$\tilde G_E = G_E + \delta \tilde G_E$, where $G_M$ and $G_E$ are the
usual proton magnetic and electric form factors, which are functions of
$Q^2$ only and are defined from matrix elements of the electromagnetic
current. The amplitudes $ \tilde{F}_{3} = \delta \tilde F_3$, $\delta
\tilde{G}_{M}$, and $\delta \tilde G_E$, originate from processes
involving the exchange of at least two photons, and are of order $e^2$
(relative to the factor \( e^{2} \) in Eq.~(\ref{eq:tmatrix})).

The unpolarized cross section is
\ba													\label{eq:reduced}
{d\sigma \over d\Omega_{Lab}} = 
				{ \tau \sigma_R  \over \epsilon(1+\tau)}
				{d\sigma_{NS} \over d\Omega_{Lab}}  \ ,
\ea

\noindent  where the ``no structure'' cross section is
\ba
{d\sigma_{NS} \over d\Omega_{Lab}} 
		= {4\alpha^2 \cos^2 (\theta / 2) \over Q^4} \,
		{E^{\prime 3} \over E}  \ ,
\ea

\noindent and $E$ and $E'$ are the incoming and outgoing electron Lab energies.  Other quantities are defined after Eq.~(\ref{eq:polratio}).  The reduced cross section including the two-photon exchange correction becomes~\cite{GV03}
\be
\sigma_R 
= G_M^2 + \frac{\varepsilon}{\tau}  G_E^2   \,  
+ 2 \, G_M {\cal R} 
\left(\delta \tilde G_M + \varepsilon \frac{\nu}{M^2} \tilde F_3 \right) 
+ 2 \frac{\varepsilon}{\tau} G_E \, 
{\cal R} \left(\delta \tilde G_E + \frac{\nu}{M^2} \tilde F_3 \right) 
+  {\mathcal{O}}(e^4) ,
\label{eq:crossen} 
\ee
where \( {\cal R} \) stands for the real part.

The general expressions for the double polarization observables, including two-photon exchange, are~\cite{GV03}:
\begin{eqnarray}
							\label{eq:pt} 
P_s &=& A_s =										\\[1.2ex]
=&&	\hskip -8mm
	-\,\sqrt{\frac{2\varepsilon (1 -\varepsilon)}{\tau}} \,\frac{(2h_e)}{\sigma_R} \,
	\left\{G_E G_M 
	+ G_E \, {\cal R} \left(\delta \tilde G_M \right) 
	+ G_M \, {\cal R} \left(\delta \tilde G_E + \frac{\nu}{M^2} \tilde F_3
	\right) 
	+  {\mathcal{O}}(e^4) \right\}  ,
										\nonumber
																\\[1.75ex]
P_l &=& - A_l 	 = \sqrt{1 - \varepsilon^2}  \,\frac{(2h_e)}{\sigma_R} \,
	\left\{G_M^2 
	+ 2 \, G_M \, {\cal R} \left(\delta \tilde G_M 
	+ \frac{\varepsilon}{1 + \varepsilon} \frac{\nu}{M^2} \tilde F_3 \right) 
	+  {\mathcal{O}}(e^4) \right\}, 
													     \nonumber 
\end{eqnarray}

\noindent  where $h_e = \pm 1/2$ is the helicity of the electron and we assumed $m_e = 0$.  The polarizations are related to the analyzing powers $A_s$ or $A_l$,  by time-reversal invariance.  That the polarization $P_l$ is unity in the backward direction, $\varepsilon =0$, follows generally from lepton helicity conservation and angular momentum conservation.

%%%%%%%%%%%%%%%%%%%%%%%%%%%%%%%%%%%%%%%%%%%%%%%%%%%%%%%%%%%%%%%%%%%%%%%%%%%%%%%%

\subsection{Two-photon exchange at the quark level}

\label{sec:3}

%%%%%%%%%%%%%%%%%%%%%%%%%%%%%%%%%%%%%%%%%%%%%%%%%%%%%%%%%%%%%%%%%%%%%%%%%%%%%%%%

\begin{figure}[ht]
\centerline{ \includegraphics[height=3.0cm]{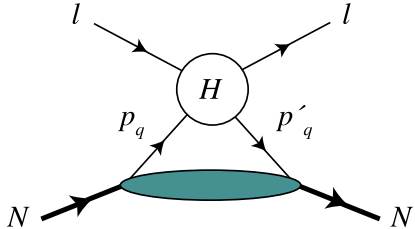} }
\caption{Handbag approximation for the elastic lepton-nucleon
scattering at large momentum transfers. In the partonic scattering process 
(indicated by $H$), the lepton scatters from quarks in the nucleon, 
with momenta $p_q$ and $p'_q$. 
The lower blob represents the GPDs of the nucleon.}
\label{fig:handbag}
\end{figure}

In order to estimate the two-photon exchange contribution 
to $\tilde G_M$, $\tilde F_2$ and $\tilde F_3$ at large momentum
transfers, we will consider a partonic calculation illustrated in
Fig.~\ref{fig:handbag}. 
To begin, we calculate the subprocess on a quark, denoted 
by the scattering amplitude $H$ in Fig.~\ref{fig:handbag}. 
Subsequently, we shall embed the quarks in the proton as 
described through the nucleon's generalized parton distributions (GPDs).

%%%%%%%%%%%%%%%%%%%%%%%%%%%%%%%%%%%%%%%%%%%%%%
%%%%%%%%%%%%%%%%%%%%%%%%%%%%%%%%%%%%%%%%%%%%%%%

\indent
Elastic lepton-quark scattering,
\begin{equation}
\label{eq:eqscatt}
l(k)+q(p_q)\rightarrow l(k')+q(p'_q) \ ,
\end{equation}
is described by two independent kinematical invariants, $\hat s \equiv
(k + p_q)^2$ and $Q^2 = -t = -(k - k')^2$. We also introduce the
crossing variable $\hat u \equiv (k - p'_q)^2$, which satisfies $\hat s
+ \hat u = Q^2$. The $T$-matrix for the two-photon part of the
electron-quark scattering can be written as
\begin{eqnarray}
\label{eq:tmatrixhard}
H_{h, \, \lambda} \,&=&\, 
\frac{(e \, e_q)^2}{Q^{2}} \, \bar{u}(k', h)\gamma _{\mu }u(k, h) \,\cdot \, 
\bar{u} (p'_q, \lambda) \left( \, \tilde{f}_{1} \, \gamma ^{\mu }
\, +\, \tilde{f}_{3} \, \gamma \cdot K \, P_q^{\mu } \, \right) u(p_q, \lambda),
\end{eqnarray}
with $P_q \equiv (p_q + p'_q) / 2$, 
where $e_q$ is the fractional quark charge (for a flavor $q$), 
and where $u(p_q, \lambda)$ and $u(p'_q, \lambda)$ are the quark
spinors with quark helicity $\lambda = \pm 1/2$, which is
conserved in the scattering process for massless quarks.  Quark helicity conservation leads to the absence of any analog of $\tilde{F}_2$ 
in the general expansion of Eq.~(\ref{eq:tmatrix}).

\begin{figure}
\vskip 6mm
\centerline{ \includegraphics[height=23mm]{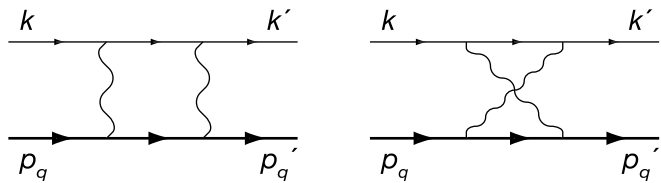} }

\caption{Direct and crossed box diagrams to describe the 
two-photon exchange contribution to the lepton-quark scattering
process, corresponding with the blob denoted by $H$ in
Fig.~\ref{fig:handbag}.}
\label{fig:hardbox}
\end{figure}

The partonic scattering helicity amplitudes $H_{h, \lambda}$ of
Eq.~(\ref{eq:tmatrixhard}) at order $O(e^4)$ are given by the
two-photon exchange direct and crossed box diagrams of
Fig.~\ref{fig:hardbox}.  
%
%%%%%%%%%%%%%%%%%%%%%%%%%%%%%%%%%%%%%%%%%%%%%%%
%%%%%%%%%%%%%%%%%%%%%%%%%%%%%%%%%%%%%%%%%%%%%%%
%
The two-photon exchange contribution to the 
elastic electron-scattering off spin 1/2 Dirac particles was
first calculated in Ref.~\cite{Nie71}, which was verified explicitly for the work reported in~\cite{YCC04,ABCCV05}.   The amplitude $\tilde f_1$, but not $\tilde f_3$, has an infrared (IR) divergence, which we isolate into a soft part, {\it i.e.},  
$\tilde{f}_1 = \tilde{f}_1^{soft} + \tilde{f}_1^{hard}$. 
The soft part corresponds with the situation where 
one of the photons in Fig.~\ref{fig:hardbox} carries zero four-momentum, 
and is obtained by replacing the other photon's four-momentum by 
$q$ in both numerator and denominator of the loop integral~\cite{grammer}. This yields,
\begin{eqnarray}
{\cal R}\left( \tilde{f}_1^{soft} \right)
&=&\frac{e^2}{4 \pi^2}\,
\left\{ \ln \left( \frac{\lambda^2}{\sqrt{- \hat s \hat u}} \right) 
\ln \left| \frac{\hat s}{\hat u} \right| + \frac{\pi^2}{2} \right\}, 
\label{eq:f1soft}
\end{eqnarray}
where $\lambda$ is 
an infinitesimal photon mass controlling the IR divergence.  The remaining $\tilde f_i$ can be found in~\cite{YCC04,ABCCV05}.

The full electron-quark elastic cross section, using Eq.~(\ref{eq:crossen}), is
\begin{eqnarray}
d \sigma \,&=&\, d \sigma_{1 \gamma} \, \left[ 
1 + 2 \, {\cal R} \left( \tilde{f}_1 \right)_{2 \gamma} + 
\varepsilon \, \frac{\hat s - \hat u}{4} \, 
 2 \, {\cal R} \left( \tilde{f}_3 \right)_{2 \gamma} \right] 
\equiv d \sigma_{1 \gamma} \, \left( 1 + \delta_{2 \gamma} \right) ,  
\label{eq:crosseq}
\end{eqnarray}
where $d \sigma_{1 \gamma}$ is the cross section in the one-photon 
exchange approximation, 
$\varepsilon = - 2 \, \hat s \, \hat u \, / \, (\hat s^2 + \hat u^2)$ 
in the massless limit, and one can easily obtain 
$\delta_{2 \gamma}$ from the $\tilde f_i$.

%%%%%%%%%%%%%%%%%%%%%%%%%%%%%%%%%%%%%%%%%%%%%%%%%%%%%%%%%%%%%%%%%%%%%%%%%%%%%%

\subsection{Calculation using generalized parton distributions}

\label{sec:4}

%%%%%%%%%%%%%%%%%%%%%%%%%%%%%%%%%%%%%%%%%%%%%%%%%%%%%%%%%%%%%%%%%%%%%%%%%%%%%%

Having calculated the partonic subprocess, 
we next discuss how to embed the quarks in the nucleon. 
We begin by discussing the soft contributions. The handbag diagrams discussed so far have both photons coupled to the same quark.  There are also 
contributions from processes where the photons interact with different quarks. 
One can show that the IR contributions from these processes, which are 
proportional to the products of the charges of the interacting quarks, 
added to the soft contributions from the handbag diagrams  
give the same result as the soft contributions calculated with just 
a nucleon intermediate state~\cite{Brodsky:1968ea}. Thus the low energy theorem for Compton scattering is satisfied.  In the handbag approximation, the hard parts which appear when the photons couple to different quarks, the so-called cat's ears diagrams, are neglected.

For the real parts, the IR divergence 
arising from the direct and crossed box diagrams, at the nucleon level,
is cancelled when adding the bremsstrahlung  
contribution  from the interference  of diagrams where a soft photon is emitted from the electron and from the proton. 
This provides a radiative correction term from the soft part of the boxes plus electron-proton bremsstrahlung 
which added to the lowest order term may be written as
\begin{eqnarray}
\label{eq:crosssoft}
\sigma_{soft} = \sigma_{1 \gamma} \, 
\left( 1 + \delta_{2 \gamma, \, soft} + \delta_{brems}^{e p} \right), 
\end{eqnarray}
where $\sigma_{1 \gamma}$ is the one-photon exchange cross 
section. In Eq.~(\ref{eq:crosssoft}), 
the soft-photon contribution due to the nucleon box diagram is given by
\begin{eqnarray}							
&&\delta_{2 \gamma , \, soft} 
= \frac{e^2}{2 \pi^2} \left\{ \ln \left( 
\frac{\lambda^2}{\sqrt{(s - M^2) |u - M^2|}}  \right) \, 
\ln \left| \frac{s - M^2}{u - M^2} \right| \right. 
													\label{eq:delta2gsoft}
													\\  \nonumber 
&&- \left. L\left( \frac{s - M^2}{s} \right) 
- \frac{1}{2} \ln^2\left( \frac{s - M^2}{s}\right) 
+ {\cal R} \left[ L\left( \frac{u - M^2}{u} \right) \right] 
+ \frac{1}{2} \ln^2\left( \frac{u - M^2}{u}\right) + \frac{\pi^2}{2} 
\right\} , 
\end{eqnarray}
where  $L$ is the Spence function defined by
%%% $E_e$ ($E_e^{\, '}$) are the initial (final) electron {\it lab} 
%%% energies, and
%
\begin{equation}
L(z) = - \int_0^z \, dt \, \frac{\ln(1 - t)}{t} \ .
\end{equation}
The bremsstrahlung contribution where a 
soft photon is emitted from an electron and proton line 
({\it i.e.}, by cutting one of the (soft) photon lines in Fig.~\ref{fig:hardbox})  
was calculated in Ref.~\cite{MT00}, which we verified explicitly, and is for the case that the outgoing electron is detected,
\begin{eqnarray}
\delta_{brems}^{e p} 
&=& \frac{e^2}{2 \pi^2} 
\bigg\{ \ln \left( \frac{4 \, (\Delta E)^2 \, (s-M^2)^2}
{\lambda^2 \, y \, ( u-M^2 )^2} \right) \, 
\ln \left( \frac{ s-M^2 }{ M^2-u } \right)  
						\nonumber \\
&& \qquad + \  L\left(1 - \frac{1}{y} \, \frac{ s-M^2 }{ M^2-u } \right)
- L\left(1 - \frac{1}{y} \, \frac{ M^2-u }{ s-M^2 } \right)
\bigg\} , 
\label{eq:deltabremsep}
\end{eqnarray}
where  $\Delta E \equiv E_e^{\, ' el} - E_e^{\, '}$ 
is the difference of the measured outgoing electron {\it lab} energy 
($E_e^{\, '}$) from its elastic value ($E_e^{\, ' el}$), and
$y \equiv (\sqrt{\tau} + \sqrt{1 + \tau})^2$.
One indeed verifies that the sum of 
Eqs.~(\ref{eq:delta2gsoft},\ref{eq:deltabremsep}) is IR finite.
When comparing with elastic $ep$ cross section data, which are 
usually radiatively corrected using the procedure of Mo and Tsai, Ref.~\cite{MoTsai68}, 
we have to consider only the difference of our 
$\delta_{2 \gamma, \, soft} + \delta_{brems}^{ep}$ relative to the ${\cal O}(Z^2)$ part, in their notation, of the radiative correction in \cite{MoTsai68}. 
Except for the $\pi^2/2$ term in Eq.~(\ref{eq:delta2gsoft}), this difference was found to be below $10^{-3}$ for all kinematics considered in Fig.~\ref{fig:cross}.

After some algebra, one obtains 
the hard $2 \gamma$ exchange contributions to 
$\delta \tilde G_M$, $\delta \tilde G_E$, and $\tilde F_3$ as,
\begin{eqnarray}
\hspace{-0.4cm}
\delta \tilde{G}_M^{hard} &=& C, 
\label{eq:GMhandbag}\\
\hspace{-0.4cm}
\delta \tilde{G}_E^{hard} &=& 
- \left( \frac{1 + \varepsilon}{2 \varepsilon} \right)\, ( A - C ) 
+ \sqrt{\frac{1 + \varepsilon}{2 \varepsilon}} \, B , 
\label{eq:GEhandbag} \\
\hspace{-0.4cm}
\tilde F_3 &=& \frac{M^2}{\nu}
\left( \frac{1 + \varepsilon}{2 \varepsilon} \right)\, (A - C ),
\label{eq:F3handbag}
\end{eqnarray}
with
\begin{eqnarray}
A &\equiv& \int_{-1}^1 \frac{dx}{x}     
\frac{\left[(\hat s - \hat u) \tilde{f}_1^{hard} -
\hat s \hat u \tilde{f}_3 \right]}{(s - u)} 
\sum_q e_q^2 \, \left( H^q + E^q \right), 
												\nonumber \\
B &\equiv& \int_{-1}^1 \frac{dx}{x}     
\frac{\left[(\hat s - \hat u) \tilde{f}_1^{hard} - 
\hat s \hat u \tilde{f}_3 \right]}{(s - u)} 
\sum_q e_q^2 \, \left( H^q - \tau E^q \right), 
												\nonumber \\
C &\equiv& \int_{-1}^1 \frac{dx}{x} \, \tilde{f}_1^{hard} \, 
\mathrm{sgm}(x) \, \sum_q e_q^2 \, \tilde H^q.
												\label{eq:ABC}
\end{eqnarray}

The functions $H^q$, $E^q$, and $\tilde H^q$ are the generalized parton distributions, which describe removing a quark of a certain momentum from a hadron, and replacing it with a quark of another momentum (or, even, in a more general case, with one of a different flavor).   One can see from Fig.~\ref{fig:handbag} that these are just what we need to knit the amplitudes for electron-quark scattering into an amplitude involving the proton. There is a fourth GPD, $\tilde E^q$, that does not enter the two-photon exchange expressions.  The GPDs can be measured in deeply virtual or wide-angle Compton scattering, and have been reviewed in~\cite{Ji:1998pc,GPV01,Diehl:2003ny,Ji:2004gf} and elsewhere.  The GPD models we use here are detailed in~\cite{ABCCV05, guidal}; so also~\cite{diehl}.

From the integrals $A$, $B$, and $C$, and the usual form factors,
we can directly construct the observables. 
The cross section is
\be
\sigma_R = \sigma_{R,soft} + \sigma_{R,hard} \,,
\ee
where
\ba
\sigma_{R,hard} 
	= (1 + \varepsilon) \, G_M \, {\cal R} \left( A \right)  
	+ \sqrt{2\, \varepsilon \, (1 + \varepsilon)} \frac{1}{\tau} \, 
			G_E \, {\cal R} \left( B \right)
	+ (1 - \varepsilon) \, G_M \, {\cal R} \left( C \right) \,.
\ea
From Eqs.~(\ref{eq:crosssoft}) to~(\ref{eq:deltabremsep}) and the discussion surrounding them, we learned that to a good approximation the result for the soft part can be written as
\ba
\sigma_{R,soft} = \sigma_{R, 1\gamma} \left( 1 + \pi\alpha + \delta^{MT} \right) 
																		\,,
\ea
where $\delta^{MT}$ is the correction given in Ref.~\cite{MoTsai68}.  Since the data is very commonly corrected using~\cite{MoTsai68}, let us define
$\sigma_R^{\ MT\ corr} \equiv \sigma_R / (1+\delta^{MT})$.  Then an accurate relationship between the data with Mo-Tsai corrections already included and the form factors is
\ba
\sigma_R^{\ MT\ corr} 
	= \left( G_M^2 + \frac{\varepsilon}{\tau} G_E^2 \right)
		( 1 + \pi\alpha )
	+ \sigma_{R,hard} \,,
\ea
where the left hand side is what experimenters often quote as radiatively corrected data.  Since the Mo-Tsai corrections are so commonly made in experimental papers before reporting the data, the ``$MT\ corr$'' superscript will be understood rather than explicit when we show cross section plots below.  The extra terms on the right-hand-side come from new two-photon exchange corrections.  The reader may marginally improve the expression by including with the $(1+\pi\alpha)$ factor the circa 0.1\% difference between our actual soft results and those of~\cite{MoTsai68}.   Finally, before discussing polarization, the fact that a $\pi^2/2$ term, or $(\pi\alpha)$ term after multiplying in the overall factors, sits in the soft corrections has to do with the specific criterion we used, that of Ref.~\cite{grammer}, to separate the soft from hard parts.  The term cannot be eliminated; with a different criterion, however, that term can move into the hard part.

The double polarization observables of Eqs.~(\ref{eq:pt}) 
are given by
\begin{eqnarray}
P_s &=& -  
\,\sqrt{\frac{2\varepsilon (1 - \varepsilon)}{\tau}} \,\frac{1}{\sigma_R} \,
\left\{G_E G_M 
+ G_E \, {\cal R} \left( C \right) 
+ G_M \, \sqrt{\frac{1 + \varepsilon}{2 \, \varepsilon}} \, 
{\cal R} \left( B \right) 
+  {\mathcal{O}}(e^4) \right\}, 
\\
P_l &=& 
\sqrt{1 - \varepsilon^2}  \,\frac{1}{\sigma_R} \,
\left\{G_M^2 
+ \, G_M \, {\cal R} \left( A + C \right) 
+  {\mathcal{O}}(e^4) \right\}  .
\end{eqnarray}

%%%%%%%%%%%%%%%%%%%%%%%%%%%%%%%%%%%%%%%%%%%%%%%%%%%%%%%%%%%%%%%%%%%%%%%%%%%%%

\subsection{Results}

\label{sec:5}

%%%%%%%%%%%%%%%%%%%%%%%%%%%%%%%%%%%%%%%%%%%%%%%%%%%%%%%%%%%%%%%%%%%%%%%%%%%%%

%%%%%%%%%%%%%%%%%%%%%%%%%%%%%%%%%%%%%%%%%%%%%%%%%%%%%%%%%%%%%%%%%%%%%%%%%%%%%

\subsubsection{Cross section}

%%%%%%%%%%%%%%%%%%%%%%%%%%%%%%%%%%%%%%%%%%%%%%%%%%%%%%%%%%%%%%%%%%%%%%%%%%%%%

%%%%%%%%%%%%%%%%%%%%%%%%%%%%%%%%%%%%%%%%%%%%%%
\begin{figure}[t]

\centerline{
\includegraphics[width=5.5cm]{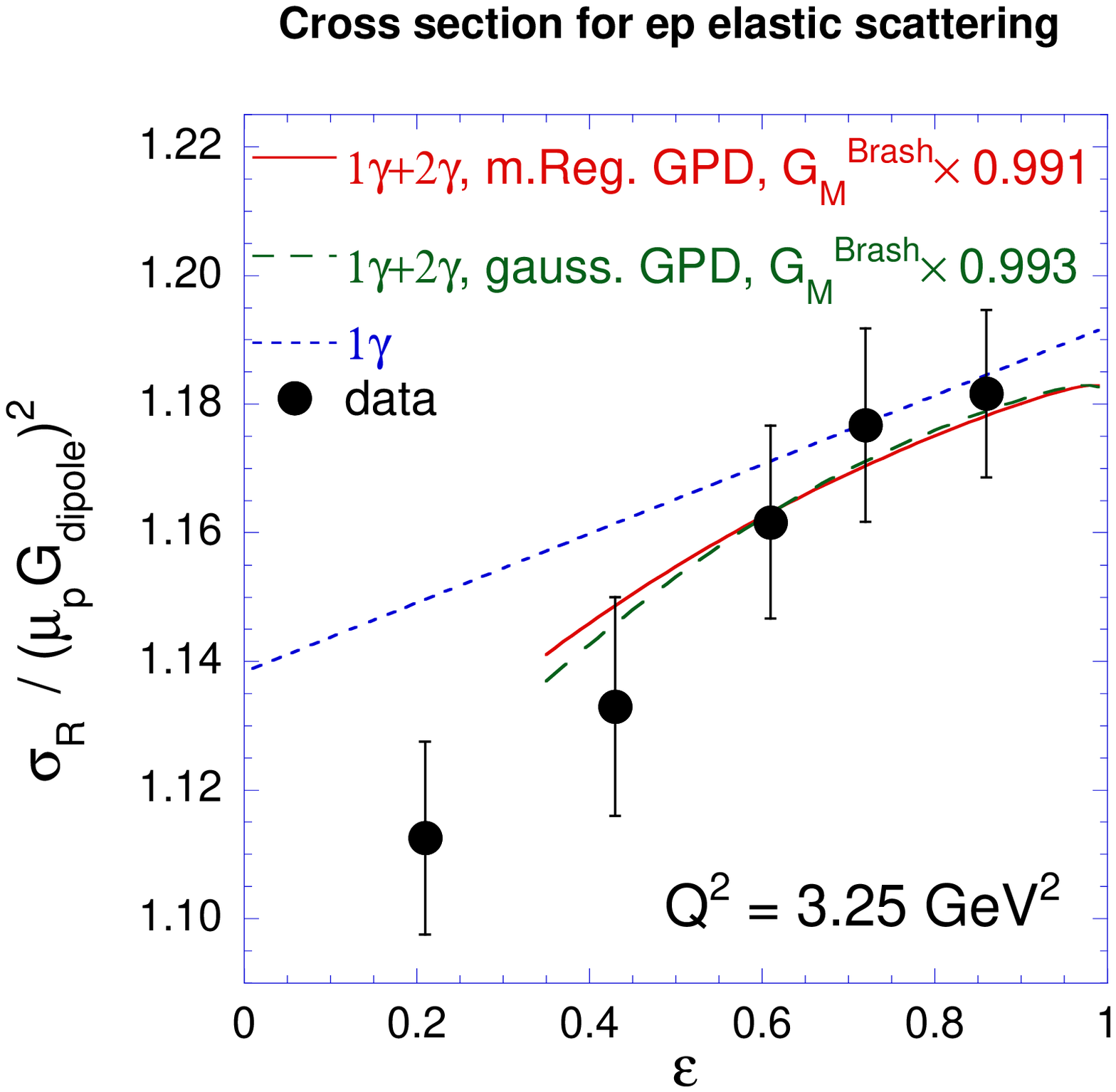} \qquad
\includegraphics[width=5.5cm]{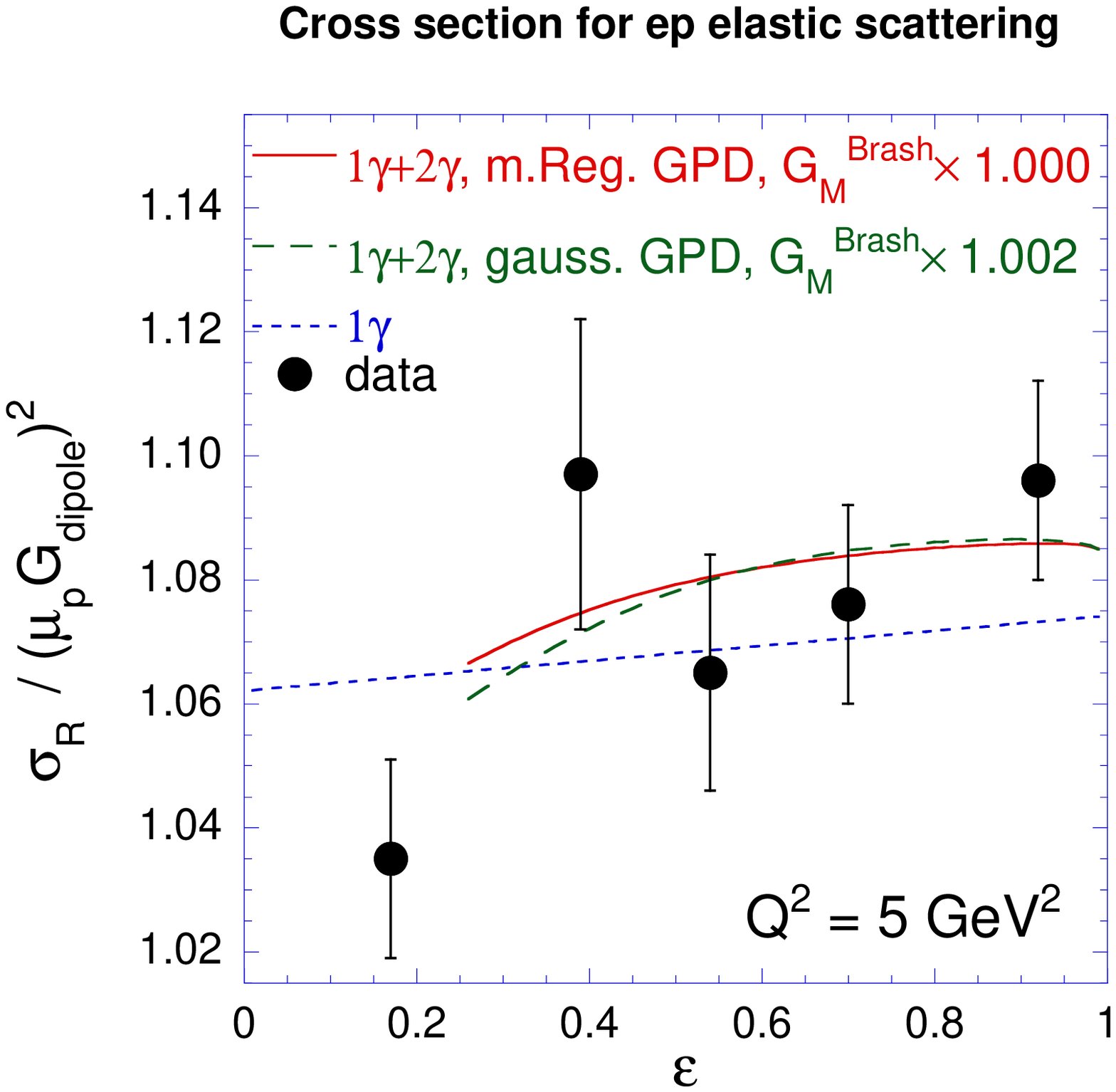} 
}

\caption{Rosenbluth plots for elastic $e p $ scattering: 
$\sigma_R$ divided by $(\mu_p G_D)^2$, 
with $G_D = (1 + Q^2 / 0.71)^{-2}$.  
Dotted curves: Born approximation using $G_{E } / G_{M }$ from 
polarization data~\cite{Jones00,Gayou02}.  
Solid curves: full calculation using the modified Regge GPD, 
for the kinematical range $-u > M^2$. 
Dashed curves: same as solid curves but using the gaussian GPD.   
The data are from Ref.~\cite{Slac94}; the Figure and calculation are from Ref.~\cite{ABCCV05}.}

\label{fig:cross}

\end{figure}
%%%%%%%%%%%%%%%%%%%%%%%%%%%%%%%%%%%%%%%%%%%%%%

Figure~\ref{fig:cross} shows the reduced differential cross section for electron-proton scattering $\sigma_R$, for two values of $Q^2$.  There are three items on each graph.  One is the data.  The next is the straight line, which is the result of the $1$-$\gamma$ exchange calculation using $G_E/G_M$ taken from the polarization data~\cite{Gayou02}, with a reasonable and commonly used choice for $G_M$~\cite{Brash:2001qq}.  The slope is too flat to fit the data, reflecting the conflict between the polarization measurements and the Rosenbluth measurements with the hard $2$-$\gamma$ corrections.  Third are the slightly curved lines, showing the results of the $2$-$\gamma$ corrections while still using the $G_E/G_M$ ratio from the polarization data.  Results are shown for two different model GPDs, described in~\cite{YCC04,ABCCV05,guidal}; they do not greatly differ.  (The renormalization of $G_M$ that we have allowed does not affect the slope.)  One sees that the hard $2$-$\gamma$ corrections steepen the average slope and improve the agreement with the data.  It is also important to note the non-linearity in the Rosenbluth plot, which can be checked with a more precise experiment.

\begin{figure}[t]%[htbp]

\centerline{ \includegraphics[width=55mm]{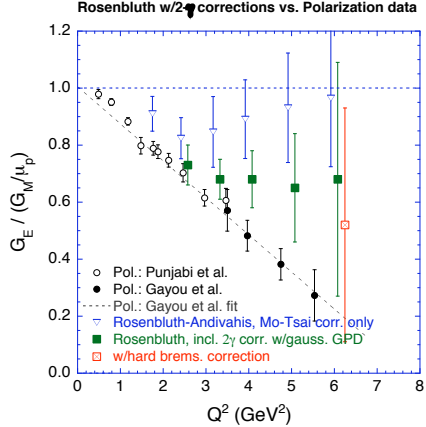} }
\caption{Determinations of the proton $G_E/G_M$ ratio.  The polarization data is from Gayou {\it et al.}~\cite{Gayou02} and Punjabi {\it et al.}~\cite{Punjabi:2005wq}, and the Rosenbluth data is from Andivahis {\it et al.}~\cite{Slac94}, which include only the well-known Mo-Tsai corrections. Our Rosenbluth $G_E/G_M$ include the two-photon corrections, and for one point also a hard bremsstrahlung correction, still using Andivahis {\it et al.}~data.  Some of our points for the Rosenbluth results are slightly offset horizontally for clarity.}
\label{newcomparison}

\end{figure}

Fig.~\ref{newcomparison} presents the $2$-$\gamma$ results in a different way.  The plot shows the extracted $G_{E} / G_{M}$ vs. $Q^2$.  One set of data points, falling linearly with $Q^2$, is from the polarization experiments.  Another set of data points, roughly constant in $Q^2$ and plotted with inverted triangles, is from Rosenbluth data analyzed using only the Mo-Tsai~\cite{MoTsai68} radiative corrections.  The solid squares show the best fit $G_{E} / G_{M}$ from the Ref.~\cite{Slac94} data when analyzed including the hard $2$-$\gamma$ corrections.  One sees that for $Q^2$ in the 2--3 GeV$^2$ range, the $G_E/G_M$ extracted using the Rosenbluth method including the $2$-$\gamma$ corrections agree well with the polarization transfer results;  At higher $Q^2$, there is at least partial reconciliation between the two methods.

%%%%%%%%%%%%%%%%%%%%%%%%%%%%%%%
%%%%%%%%%%%%%%%%%%%%%%%%%%%%%%%

There is one more point on Fig.~\ref{newcomparison}, which shows the result of also including some hard bremsstrahlung corrections, which will be discussed below.

%%%%%%%%%%%%%%%%%%%%%%%%%%%%%%%%%%%%%%%%%%%%%%%%%%%%%%%%%%%%%%%%%%%%%%%%%%%%%

\subsubsection{Polarization transfers}

%%%%%%%%%%%%%%%%%%%%%%%%%%%%%%%%%%%%%%%%%%%%%%%%%%%%%%%%%%%%%%%%%%%%%%%%%%%%%

\begin{figure}[t]

\centerline{
\includegraphics[width=5.5cm]{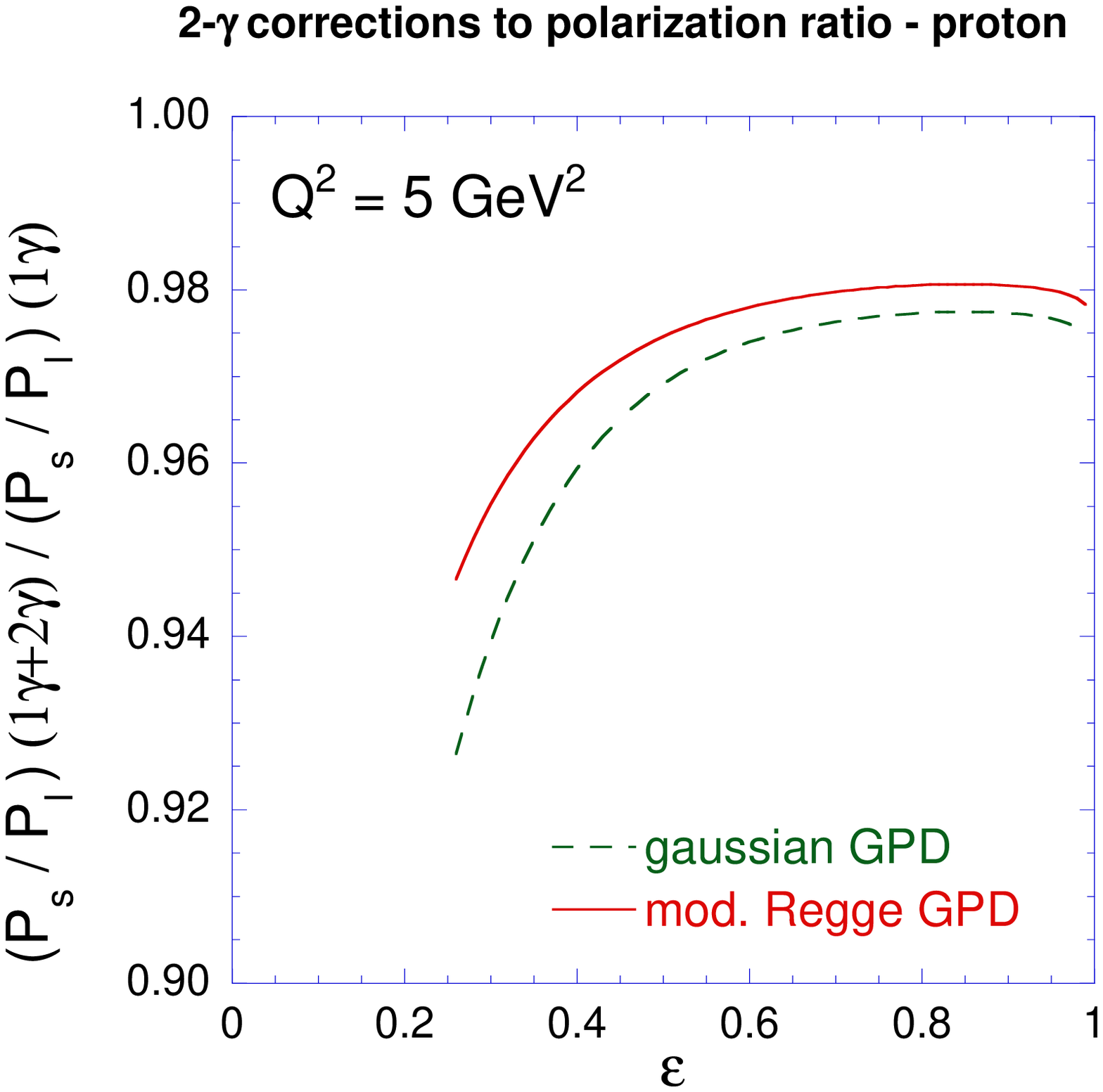} \qquad
\includegraphics[width=5.5cm]{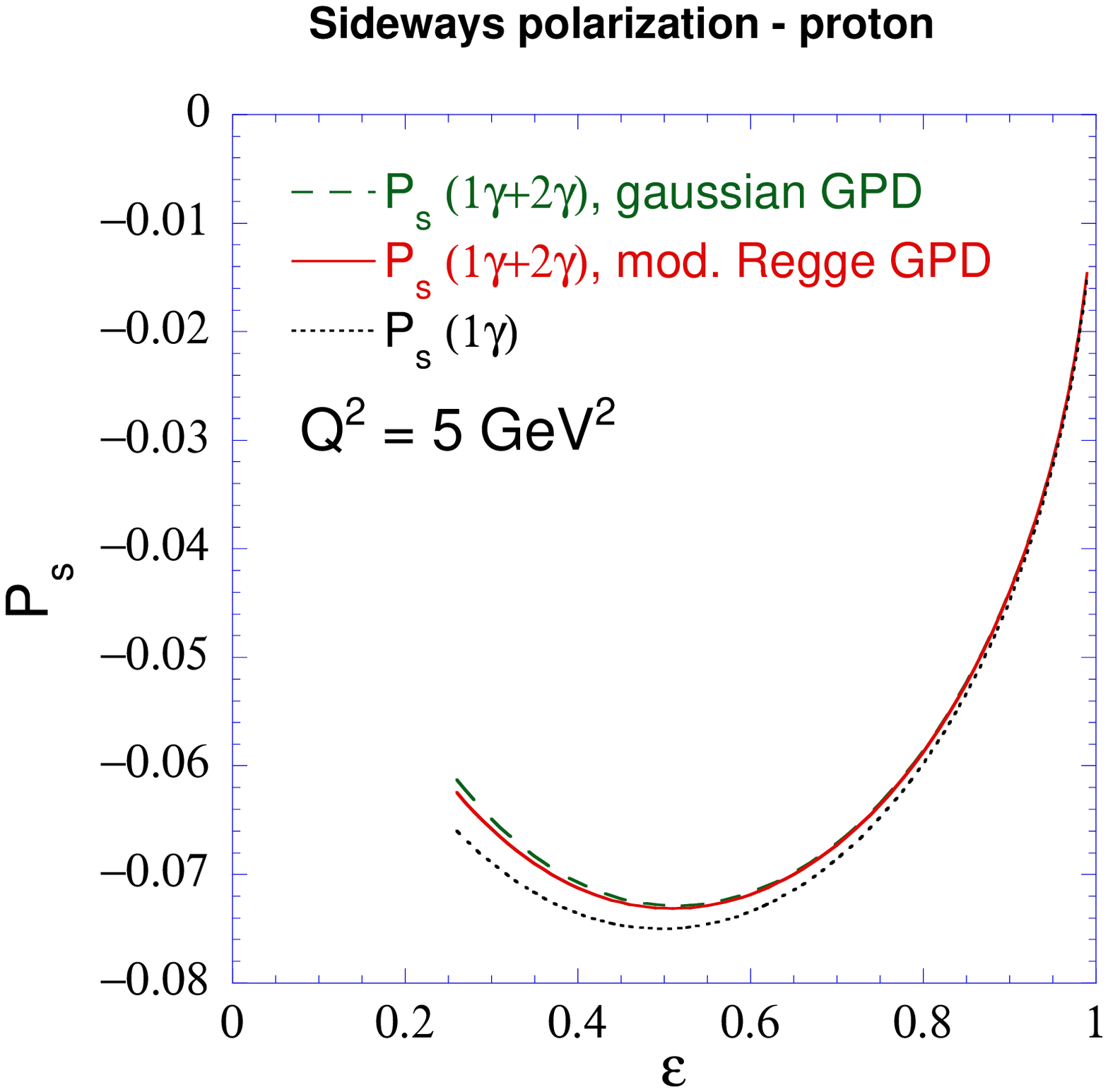}
}

\caption{Polarization $P_s/P_l$ and $P_s$ compared to the $1 \gamma$ exchange results for $e p$ scattering at $Q^2$ = 5 GeV$^2$. 
The solid and dashed curves show the $2$-$\gamma$ exchange correction using the 
GPD calculation, for the kinematical range where both $s, -u > M^2$.  The dotted curves on the right shows the Born approximation ($1 \gamma$ exchange) result. }
\label{fig:plpt}
\end{figure}

The $2$-$\gamma$ corrections do not impact the polarizations measurements as strongly as the Rosenbluth measurements.  The left panel of Fig.~\ref{fig:plpt} shows the correction to the $P_s/P_l$ ratio from the hard $2$-$\gamma$ exchange.  Most of the effect is on $P_s$, shown separately in the right panel of the same Figure; the effect on $P_l$ is too small to show on Figures like these.  The present polarization experiments have $\varepsilon \approx 0.7$.  The $2$-$\gamma$ corrections induce an $\varepsilon$-dependence that could be seen in a precise experiment ~\cite{charles}.

%%%%%%%%%%%%%%%%%%%%%%%%%%%%%%%%%%%%%%%%%%%%%%%%
%%%%%%%%%%%%%%%%%%%%%%%%%%%%%%%%%%%%%%%%%%%%%%

%%%%%%%%%%%%%%%%%%%%%%%%%%%%%%%%%%%%%%%%%%%%%%%%%%%%%%%%%%%%%%%%%%%%%%%%%%%%%

\subsubsection{Positron-proton vs. electron-proton}

%%%%%%%%%%%%%%%%%%%%%%%%%%%%%%%%%%%%%%%%%%%%%%%%%%%%%%%%%%%%%%%%%%%%%%%%%%%%% 

\begin{figure}[t]
\centerline{ \includegraphics[width=58 mm]{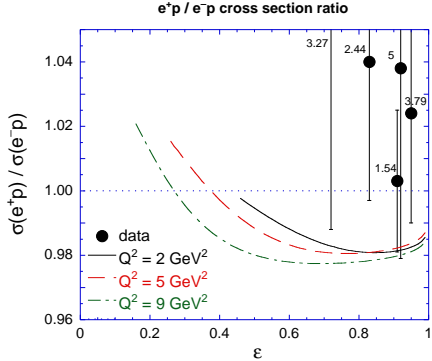} }
\caption{Ratio of $e^+ / e^-$ elastic cross sections on the proton.    
The GPD calculations of the $2 \gamma$ correction are for $Q^2$ of 2, 5, and 9 GeV$^2$, 
for the kinematical range with $-u$ above $M^2$. 
Also shown are all known data, from~\cite{Mar68}, with $Q^2$ above 1.5 GeV$^2$ (the missing central value is at 1.111).    The numbers near the data give $Q^2$ for that point in GeV$^2$.
}

\label{fig:positron}
\end{figure}

Positron-proton and electron-proton scattering have the opposite sign for the two-photon corrections 
relative to the one-photon terms.  Hence one expects $e^+ p$ and $e^- p$ elastic scattering to differ 
by a few percent.  Figure~\ref{fig:positron} shows our results for three different $Q^2$ values.  
These curves are obtained by adding our two-photon box calculation, minus the corresponding part of the soft 
only calculation in~\cite{MoTsai68}, to the one-photon calculations; hence, they are meant to be compared to 
data where the corrections given in~\cite{MoTsai68} have already been made.  Each curve is based on the gaussian 
GPD and is cut off at low $\varepsilon$ when $-u = M^2$.  Early data from SLAC are available~\cite{Mar68}; more 
precise data are anticipated from JLab~\cite{brooks}.  (Ref.~\cite{Mar68} used the Meister-Yennie~\cite{oldyennie} 
soft corrections rather than those of Mo and Tsai.  We have checked that for these kinematics the difference between 
them is  smaller than $ 0.1\%$,  which is negligible compared to the size of the error bars.)

%%%%%%%%%%%%%%%%%%%%%%%%%%%%%%%%%%%%%%%%%%%%%%

%%%%%%%%%%%%%%%%%%%%%%%%%%%%%%%%%%%%%%%%%%%%%%

%%%%%%%%%%%%%%%%%%%%%%%%%%%%%%%%%%%%%%%%%%%%%%%%%%%%%%%%%%%%%%%%%%%%%%%%%%%%%

\subsubsection{Results from single-baryon intermediate states}

%%%%%%%%%%%%%%%%%%%%%%%%%%%%%%%%%%%%%%%%%%%%%%%%%%%%%%%%%%%%%%%%%%%%%%%%%%%%% 

\begin{figure}[t]
\centerline{ \includegraphics[width=58 mm,angle=-90]{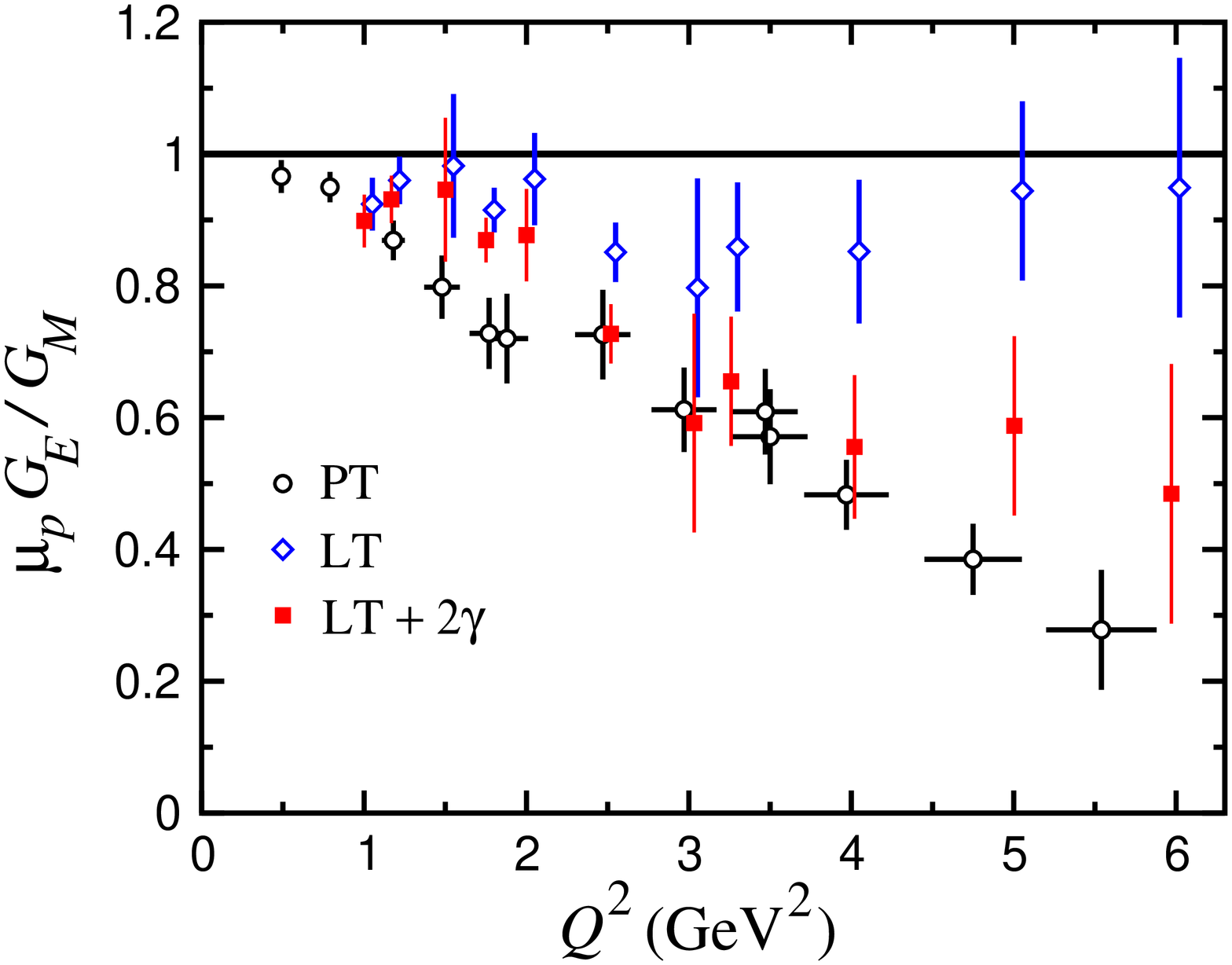} }
\caption{Extracting $G_E/G_M$ with $2\gamma$ corrections calculated using a single proton as the hadronic intermediate state.  ``PT'' is $G_E/G_M$ obtained from the polarization transfer experiments;``LT'' is $G_E/G_M$ obtained from a Rosenbluth experiment using only the Mo-Tsai radiative corrections, and ``LT+$2\gamma$'' includes the extra $2\gamma$ corrections done this way.  This figure is a based on a figure in Ref.~\cite{Blunden:2005ew}, and we thank the authors for providing it.
}

\label{fig:wally3}
\end{figure}

We have focused on a partonic view of the two-photon physics.  The results when viewing the hadronic intermediate state as a proton, or a proton plus a set of resonances, are similar~\cite{BMT03,Blunden:2005ew,Kondratyuk:2005kk,Kondratyuk:2007hc,Borisyuk:2006fh}.  The effect, in a calculation with just a proton intermediate state, of the extra $2$-$\gamma$ corrections upon extracting $G_E/G_M$ from a Rosenbluth experiment is shown in Fig.~\ref{fig:wally3}.  Further, corrections to the polarization experiments are just a few percent, in the same direction as found in the partonic evaluation, with nearly all the effect coming upon $P_s$ and not on $P_l$~\cite{Blunden:2005ew}.

%%%%%%%%%%%%%%%%%%%%%%%%%%%%%%%%%%%%%%%%%%%%%%

%%%%%%%%%%%%%%%%%%%%%%%%%%%%%%%%%%%%%%%%%%%%%%

%%%%%%%%%%%%%%%%%%%%%%%%%%%%%%%%%%%%%%%%%%%%%%%%%%%%%%%%%%%%%%%%%%%%%%%%%%%%% 

\subsection{Remarks on related topics}

\label{sec:theend}

%%%%%%%%%%%%%%%%%%%%%%%%%%%%%%%%%%%%%%%%%%%%%%%%%%%%%%%%%%%%%%%%%%%%%%%%%%%%% 

There has been a suggestion that hard bremsstrahlung may cause the difference between the Rosenbluth and polarization results~\cite{Bystritskiy:2006ju}.  Bremsstrahlung means a process where a real photon is emitted.  If the photon energy is sufficiently low, the experimenters will fail to see it and will count the reaction as elastic.  Usual bremsstrahlung calculations are for soft bremsstrahlung, where the emitted photon energy is kept only to linear order in denominators and entirely omitted in numerators.  Soft bremsstrahlung multiplies all amplitudes by the same factor and does not, for a relevant example, change the slope on a Rosenbluth plot.  If one makes no approximations in the photon energy, there can be different effects on different spin amplitudes.  Thus the claim is that emitted photons that are energetic enough to affect the spin structure of the calculation but still small enough to escape detection, give rise to the difference between the two methods of measuring $G_E/G_M$.  A contrasting numerical claim is that hard bremsstrahlung effects are noticeable and helpful in reconciling the Rosenbluth and polarization experiments, but are not decisive.  Along these lines is a result for hard bremsstrahlung at $Q^2 = 6$ GeV$^2$ from Afanasev~\cite{afanasev2005}, which has been added to the $2$-$\gamma$ results and included in Fig.~\ref{newcomparison}.

These contrasting claims clearly need adjudication, but an independent reexamination is not available as of this writing.

Electron, or muon, scattering off deuterons or larger nuclei has not been within the scope of the present review.  Larger nuclei have a factor $Z$ advantage in the relative size of the $2$$\gamma$ and $1$$\gamma$ effects, although breakup effects vitiate this advantage for elastic scattering except at low energy.  One can examine some of the work seeking evidence of $2$$\gamma$ effects in larger nuclei in~\cite{Dong:2006wm,Rekalo:1999mt,Bordes:1986km,Penarrocha:1980fx,Bernabeu:1980bi}.

Two-photon exchange effects also affect parity-violating $e$-$p$ elastic scattering via their interference with the lowest order $Z$-exchange diagram.  Ref.~\cite{Afanasev:2005ex} pointed this out, and found that the $2$$\gamma$ exchange also led to extra terms with different $\tau$ and $\varepsilon$ dependences than those known from analyses using only Born diagrams.  The calculated size of the effects, using the partonic model at $Q^2$ of several GeV$^2$, was of ${\cal O}(1\%)$.  This is below present experimental uncertainties, but parity-violating experiments with ${\mathcal O}((1/2)\%)$ uncertainties are planned.  

Arrington and Sick~\cite{Arrington:2006hm}, considering the effects that the
most recent and precise low-$Q^2$ determinations of $G_E$ and $G_M$ would have
upon parity-violating $e$-$p$ elastic scattering, 
and included the two photon correction terms that 
were pointed out in~\cite{Afanasev:2005ex}.  
The actual two-photon calculations at their $Q^2$ were 
done using single hadron intermediate states~\cite{Blunden:2005jv}.
%%%%%%%%%%%%%%%%%%%%
\section{Polarizability corrections to hydrogen hyperfine splitting}
%%%%%%%%%%%%%%%%%%%%

%%%%%%%%%%%%%%%%%%%%
\subsection{Two-photon exchange and atomic structure}
%%%%%%%%%%%%%%%%%%%%

We begin with some explanation of how this piece of atomic physics fits properly into a review of two-photon physics.  Hydrogen hyperfine splitting (hfs) in the ground state is known
to 13 significant figures in frequency units~\cite{Karshenboim:1997zu},
    \be
    E_{\rm hfs}(e^-p) = {\rm 1\ 420.405\ 751\ 766\ 7  (9)\ MHz} \,.
    \ee
This accuracy is remarkable to theorists, who are currently hopeful of obtaining a calculation 
accurate to a part per million (ppm).   To reach this goal, some improvement is needed, and the current best calculations are a few ppm away from the data.

\begin{figure}[htbp]
\begin{center}
\includegraphics[height=22mm]{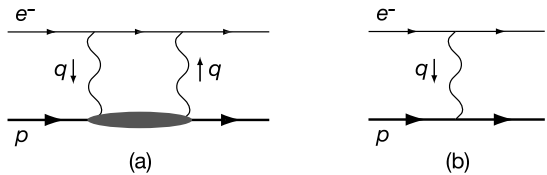}
\caption{(a) Generic two-photon exchange diagram, giving proton-structure corrections to hyperfine splitting.  (b) One-photon exchange.}
\label{twophoton}
\end{center}
\end{figure}

The main uncertainty in calculating the hfs in hydrogen comes from the hadronic, or proton structure, corrections.  The generic process that contributes is two-photon exchange, shown in Fig.~\ref{twophoton}a, which involves the proton structure because each of the exchanged photons could individually be quite energetic.

One-photon exchange, Fig.~\ref{twophoton}b, does not involve proton structure, at the accuracy needed for the present purpose.  The characteristic momentum of the electron in a hydrogen atom is of ${\mathcal O}(\alpha m_e)$, which is very low on a nuclear physics scale.  Hence the $q^2$ of an exchanged single proton is very low, and the variation of the proton form factor from its $q^2=0$ value is minimal.  One can show that keeping the electron momentum gives corrections of ${\mathcal O}(\alpha m_e/M)$ smaller than what comes from two-photon exchange.  Hence one sets the momenta of the electrons to zero.  (For information, in the one-photon exchange hfs calculation there comes a $q^2$ factor in the numerator which cancels the $1/q^2$ from the photon propagator;  then the neglect of the electron momenta is safe.)

%%%%%%%%%%%%%%%%%%
%%%%%%%%%%%%%%%%%%

%%%%%%%%%%%%%%%%%%%%
\subsection{Hyperfine splitting calculations}
%%%%%%%%%%%%%%%%%%%%

The calculated hyperfine splitting in hydrogen is~\cite{Karshenboim:1997zu,Volotka:2004zu,dupays},
\ba
    E_{\rm hfs}(e^-p) =
    \big (1+\Delta_{\rm QED} + \Delta_{\rm weak}^p+\Delta_{\rm str} \big)   \, E_F^p 	\,;
\ea
the two-photon exchange lies in the structure dependent term $\Delta_{\rm str}$.  The Fermi energy is
\be
    E_F^p=\frac{8 \alpha^3 m_r^3 }{3\pi} \mu_B\mu_p
    	=	\frac{16  \alpha^2}{3} \frac{\mu_p}{\mu_B}  
		\frac{ R_\infty }{ \left( 1 + m_e/M \right)^3 }			\,,	
\ee
where $m_r = m_e M / (M + m_e)$ is the reduced mass.  By convention, in $E_F^p$ one uses the actual magnetic moment for the proton and the Bohr magneton for the electron (note that $\mu_B$ can be used to replace the electron mass), and $R_\infty$ is the Rydberg constant in frequency units.  The second form allows optimal accuracy in evaluating $E_F^p$.  The least accurately known quantity is the ratio $\mu_p/\mu_B$, which is known to 8 figures.  Hence to the ppm level,  $E_F^p$ is known more than sufficiently well.  The QED~\cite{Eides:2000xc,Karshenboim:2005iy} and weak interaction corrections~\cite{Eides:1995sq} are well known and will not be discussed, except to mention that the QED corrections could be obtained, for the present purposes, without calculation.  They are the same as for muonium, so it is possible to obtain them to an accuracy more than adequate for the present purpose using muonium hfs data and a judicious scaling~\cite{Brodsky:2004ck,Friar:2005rs}. 

The structure dependent corrections are, in a standard treatment, divided into Zemach, recoil, and polarizability corrections,
\ba
\Delta_{\rm str} = \Delta_Z +  \Delta_R^p +  \Delta_{\rm pol}		\,.
\ea

\begin{figure}[htbp]
\begin{center}
\includegraphics[height=22mm]{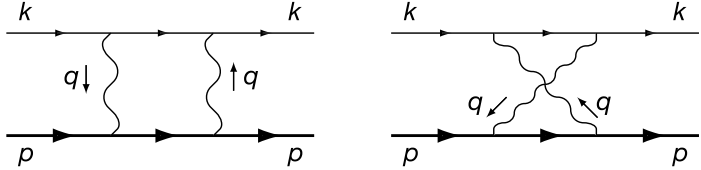}
\caption{Two-photon exchange diagrams for the ``elastic''  proton-structure corrections to hyperfine splitting.}
\label{boxes}
\end{center}
\end{figure}

The first two terms arise when the blob in Fig.~\ref{twophoton}a is just a proton, as in Fig.~\ref{boxes}, and they are together called the elastic corrections.  The electron-photon vertex is well known, and the proton-photon vertex is given by~\cite{Bodwin:1987mj,Martynenko:2004bt}
\ba
\label{vertex}
\Gamma_\mu = \gamma_\mu F_1(Q^2) + \frac{i}{2M} \sigma_{\mu\nu} q^\nu F_2(Q^2)
\ea
(for the photon with incoming momentum $q$) \underline{{\it if}} the intermediate proton is on-shell.  Of course, it is generally not.  However, one can show that the imaginary part of the diagram does come only from kinematics where the intermediate electron and proton are on-shell.  Hence, one can correctly use the above vertex to calculate the imaginary part of the diagrams, and then obtain the whole of the diagram using a dispersion relation.

%%%%%%%%%%%%%%%%%%
%%%%%%%%%%%%%%%%%%

In the nonrelativistic limit, the recoil terms are zero and the Zemach term is not.  (The nonrelativistic limit is $M \to \infty$, with $m_e$ and the proton size held fixed. Proton size information is embedded in the form factors $F_1$ and $F_2$.)  The Zemach term~\cite{Zemach} was calculated long ago and in modern form is,
\ba
\Delta_Z = \frac{8 \alpha m_r}{\pi} \int_0^\infty \frac{dQ}{Q^2} \ 
	\left[   G_E(Q^2)  \frac{ G_M(Q^2) }{ 1+ \kappa_p } - 1   \right]
		=  - 2 \alpha m_r  r_Z  \,,
\ea
the last equality defining the Zemach radius $r_Z$.  The charge and magnetic form factors are linear combinations of $F_1$ and $F_2$,
\ba
G_M &=& F_1 + F_2  \,, \nonumber \\
G_E &=& F_1 - \frac{Q^2}{4 M^2}  F_2  \,.
\ea

The recoil corrections won't be explicitly displayed in this paper because they are somewhat long (although not awful; see~\cite{Bodwin:1987mj,Martynenko:2004bt}).  An important point is that they do depend on the form factors and hence upon the proton structure.  However, their numerical value is fairly steady by present standards of accuracy when they are evaluated using different up-to-date analytic form factors based on fits to the scattering data.

One gets the polarizability corrections by summing all contributions where the intermediate hadronic state, the blob in Fig.~\ref{twophoton}a, is not a single photon.  Paralleling the elastic case, one can show that the imaginary part of this diagram comes only from configurations where the intermediate electron plus hadronic state is kinematically on-shell, {\it i.e.}, physically realizable.  Hence one can calculate the imaginary part of the box diagram if one has data on inelastic electron-proton scattering.  Then, further paralleling the elastic case, one obtains the full box diagram via a dispersion relation.

To see some detail, the lower half of Fig.~\ref{twophoton}a is the same as forward Compton scattering of off-shell photons from protons, which is given in terms of the matrix element
\begin{eqnarray}			
T_{\mu\nu}(q,p,S) = \frac{i}{2\pi M} \int d^4\xi \ e^{iq{\cdot}\xi}
	\left\langle pS \right| T \left\{ j_\mu(\xi), j_\nu(0) \right\} \left| pS \right\rangle \,,
\end{eqnarray}
where $j_\mu$ is the electromagnetic current and the states are proton states of momentum $p$ and spin 4-vector $S$.  The spin dependence is in the antisymmetric part
\begin{eqnarray}			
\label{eqn:ta}
T^A_{\mu\nu} = \frac{i}{M\nu} \, \epsilon_{\mu\nu\alpha\beta} q^\alpha 
	\left[ \left(H_1(\nu,q^2) + H_2(\nu,q^2) \right) S^\beta - H_2(\nu,q^2) 
		\frac{ S{\cdot}q \ p^\beta }{ p{\cdot}q }		\right]  \,.
\end{eqnarray}
The two structure functions $H_1$ and $H_2$ depend on $q^2$ and on the lab frame photon energy $\nu$, defined by $M \nu = p\cdot q$.

The optical theorem that relates the imaginary part of the forward Compton amplitude to the inelastic scattering cross section for off-shell photons on protons.  The relations precisely are
\ba			
{\rm Im\,} H_1(\nu,q^2) = \frac{1}{\nu} \  g_1(\nu,q^2)
			\qquad  {\rm and} \qquad 
{\rm Im\,} H_2(\nu,q^2)  = \frac{M}{\nu^2} \   g_2(\nu,q^2)   \,,
\ea
where $g_1$ and $g_2$ are functions appearing in the cross section and are measured~\cite{Anthony:2000fn,Anthony:2002hy,Fatemi:2003yh,models,deur}  at SLAC, HERMES, JLab, and elsewhere.

Using the Compton amplitude to evaluate the inelastic part  of the two-photon loops leads to
\ba
\Delta_{\rm pol} &=& \left. \frac{E_{2\gamma}}{E_F}  \right|_{\rm inel}
	= \frac{2\alpha m_e}{ ( 1+\kappa_p) \pi^3 M  }
				\\  \nonumber
&&	\times \ \int \frac{d^4Q}{(Q^4+4m_e^2 Q_0^2) Q^2} 
		\left\{ (2Q^2 + Q_0^2) H^{\rm inel}_1(iQ_0,-Q^2)
		 	- 3 Q^2 Q_0^2 H^{\rm inel}_2(iQ_0,-Q^2) \right\} \,,
\ea
where we have Wick rotated the integral so that $Q_0 = -i \nu$, $\vec Q = \vec q$, and $Q^2 \equiv Q_0^2 + \vec Q^2$.
The dispersion relations which gives $H_{1}$ is, assuming no subtraction,  
\ba
H^{\rm inel}_1(\nu_1,q^2) = \frac{1}{\pi} \int_{\nu_{th}^2}^\infty d\nu 
	\frac{ {\rm Im\,} H_1(\nu,q^2) } { \nu^2 -\nu_1^2 }  \,,
\ea
where the integral is only over the inelastic region, and a similar relation holds for $H_2$. The no-subtraction assumption will be discussed later.

Integrating what can be integrated, and neglecting $m_e$ inside the integral, yields the expression~\cite{Iddings,Drell:1966kk,DeRafael:mc,Gnaedig:qt,Faustov:yp,Nazaryan:2005zc}
\ba
    \Delta_{\rm pol}=\frac{ \alpha m_e}{2 (1+ \kappa_p) \pi M}
    (\Delta_1+\Delta_2),
\ea
where, with $\tau = \nu^2/Q^2$,			
\ba
  \Delta_1 &=& \frac{9}{4}\int_0^\infty \frac{dQ^2}{Q^2}\left\{F_2^2(Q^2) +4 M
	\int_{\nu_{th}}^\infty	 \frac{d\nu}{\nu^2} \beta(\tau)  g_1(\nu, Q^2)
		\right\}   \,,
					\\[1ex]	\nonumber
\Delta_2 &=& -12M  \int_0^\infty \frac{dQ^2}{Q^2}
	\int_{\nu_{th}}^\infty	 \frac{d\nu}{\nu^2} \beta_2(\tau)  g_2(\nu, Q^2) .
\ea
The auxiliary functions are
\begin{eqnarray}
\beta(\tau) &=& \frac{4}{9} \left[  -3\tau + 2\tau^2
    + 2(2-\tau)\sqrt{\tau(\tau+1)}  \right]  \,,
                    \nonumber \\
\beta_2(\tau) &=& 1 + 2\tau - 2 \sqrt{\tau(\tau+1)}    \,.
\end{eqnarray}

The integral for $\Delta_1$ actually requires further comment.  Only the second terms comes from the procedure just outlined; it was historically thought convenient to add the first term, and then subtract the same term from the the recoil corrections. This stratagem allows the electron mass to be taken to zero in $\Delta_1$.  The individual terms in $\Delta_1$ diverge (they would not had the electron mass been kept), but the whole is finite because of the Gerasimov-Drell-Hearn (GDH)~\cite{Gerasimov:1965et,Drell:1966jv} sum rule, 
\ba		
4 M  \int_{\nu_{th}}^\infty	 \frac{d\nu}{\nu^2}   g_1(\nu, 0) = - \kappa_p^2  \,,
\ea
coupled with the observation that the auxiliary function $\beta(\tau)$ becomes unity as we approach the real photon point.

%%%%%%%%%%%%%%%%%%
\subsection{Numerics, especially for $\Delta_{\rm pol}$}
%%%%%%%%%%%%%%%%%%

We start this numerical section with a brief discussion of the polarizability corrections.  They have some history. Considerations of $\Delta_{\rm pol}$ were begun by Iddings in 1965~\cite{Iddings}, improved by Drell and Sullivan in 1966\cite{Drell:1966kk}, and given in present notation by de Rafael in 1971~\cite{DeRafael:mc}.  But no sufficient spin-dependent data existed, so it was several decades before the formula could be evaluated to a result incompatible with zero.  In 2002, Faustov and Martynenko became the first to use $g_{1,2}$ data to obtain results inconsistent with zero~\cite{Faustov:yp}.  They got
\ba		
\Delta_{\rm pol}({\rm F\&M\ }2002)  = (1.4 \pm 0.6) {\rm\  ppm}
\ea
However, only SLAC data was available.  None of the SLAC data had $Q^2$ below $0.30$ GeV$^2$; $\Delta_1$ and $\Delta_2$ are sensitive to the behavior of the structure functions at low $Q^2$.   Also in 2002 there appeared analytic expressions for $g_{1,2}$ fit to data by Simula, Osipenko, Ricco, and Taiuti~\cite{Simula:2001iy}, which included JLab as well as SLAC data.  Simula {\it et al.}~did not integrate their results to obtain $\Delta_{\rm pol}$, but had they done so, they would have obtained 
$\Delta_{\rm pol}  = (0.4 \pm 0.6) {\rm\  ppm}$~\cite{Nazaryan:2005zc}.

More recently Faustov and Martynenko, joined by Gorbacheva,~\cite{Faustov:2006ve} have reanalyzed their result for $\Delta_{\rm pol}$, and obtained a somewhat larger value,
\ba
\Delta_{\rm pol}({\rm FGM\ }2006) = \left( 2.2 \pm 0.8 \right) {\rm\ ppm} \,.
\ea
The data underlying this result, however, still does not include the lower $Q^2$ data from JLab that will be noted immediately below.

Data for $g_1(\nu,q^2)$ is now improved thanks to the EG1 experiment at JLab, which had its first data run in 2000--2001.  A preliminary analysis of this data became available in 2005~\cite{deur}; final data is anticipated ``soon''.  The $Q^2$ range measured in this experiment went down to $0.05$ GeV$^2$.  Using analytic forms checked against the preliminary data, Ref.~\cite{Nazaryan:2005zc} has evaluated $\Delta_{\rm pol}$ and obtained
\ba
\Delta_{\rm pol}({\rm NCG\ }2006) = \left( 1.3 \pm 0.3 \right) {\rm\ ppm} \,.
\ea
This is similar to the 2002 Faustov-Martynenko result, but with a claim that the newer data allows a smaller uncertainty limit.

A list of the numerical values of the corrections compared to the experimental value of the hfs is given is Table~\ref{table:one}.  For the polarizability corrections, we used the value from Ref.~\cite{Nazaryan:2005zc} on the grounds that is was based on the most complete inelastic electron-proton scattering data.  For the Zemach term, we used the value~\cite{Friar:2003zg} based on the form factor fits of Sick~\cite{Mohr:2000ie}, because those fits emphasized the low-$Q^2$ elastic scattering data that dominates the Zemach integral.  The values for the recoil terms and weak interaction corrections have lower uncertainty limits.  We took the former come from~\cite{Brodsky:2004ck} and they are also discussed in~\cite{Volotka:2004zu};  the latter may be found in~\cite{Volotka:2004zu,Eides:1995sq}.

%%%%%%%%%%%%%
\begin{table}[htdp]
\caption{Up-to-date corrections to hydrogenic hyperfine structure.  The first line with numbers gives the ``target value'' based on the experimental data and the best evaluation of the Fermi energy (8 figures) based on known physical constants.   The corrections are listed next.  (The Zemach term includes a 1.53\% correction from higher order electronic contributions~\cite{Karshenboim:1996ew}, as well as a +0.07 ppm correction from muonic vacuum polarization and a +0.01 ppm correction from hadronic vacuum polarization~\cite{Volotka:2004zu}.)  The total of all corrections is $1.68 \pm 0.60$ ppm short of the experimental value. }
\begin{center}

\begin{tabular}{lrc}
Quantity						&	value (ppm)	& uncertainty (ppm) \\
\hline
$({E_{\rm hfs}(e^-p)}/{E_F^p})  - 1$	&	$1\ 103.48$  &	$0.01$	\\
\hline
$\Delta_{\rm QED}$				& $1\ 136.19$	&	$0.00$	\\
$\Delta_Z$ (using Friar \& Sick~\cite{Friar:2003zg})
							&	$-41.59$	&	$0.46$	\\
$\Delta_R^p$					&	$5.84$	&	$0.15$	\\
$\Delta_{\rm pol}$ (from~\cite{Nazaryan:2005zc})		
							&	$1.30$	&	$0.30$	\\
$\Delta_{\rm weak}^p$			&	$0.06$	&	 		\\
\hline
Total							&  $1101.80$	&	$0.60$	\\
Deficit						&	$1.68$	&	$0.60$     \\
\hline
\end{tabular}

\end{center}
\label{table:one}
\end{table}%
%%%%%%%%%%%%%

Thus the sum of what one may argue are the best calculated corrections falls short of the data by about 2 ppm, or about 2.8 standard deviations.  Of course, some judgement has entered the choice of numbers.  Other form factor fits to ostensibly the same data give different values for $\Delta_Z$; for example the form factors of Kelly~\cite{Mohr:2000ie} lead to $\Delta_Z = -40.93 \pm 49$ ppm.  The reader will surely notice that using the Kelly based value of $\Delta_Z$ and the latest value of $\Delta_{\rm pol}$ from Faustov, Gorbacheva, and Martynenko leads to excellent agreement of the measured hfs.

%%%%%%%%%%%%%%%%%%%%%%%%%%%
\subsection{Comments on the derivations of the formulas}
%%%%%%%%%%%%%%%%%%%%%%%%%%%

The polarizability corrections depend on theoretical results that are obtained using unsubtracted dispersion relations.  One would like to know just what this means, and since there is at least a small discrepancy between calculation and data, one would like to be able to assess the validity of such dispersion relations.  

Dispersion relations\cite{Drechsel:2002ar} involve imagining some particular real variable to be a complex one and then using the Cauchy integral formula to find the functions of that variable at a particular point in terms of an integral around the boundary of some region.  This is a useful thing to do because, at least in particular cases, we have information from other sources about the function on part of the boundary, and legitimate reasons for neglecting contributions from the rest of the boundary.   In the present case we consider the functions $H_i(\nu,q^2)$ and we ``disperse'' in $\nu$, treating $q^2$ as a constant while we do so.  Three things are needed to make the dispersion calculation work:
\begin{itemize}
\item The Cauchy formula and knowing the analytic structure (locations of poles and cuts) of the desired amplitudes. 
\item The optical theorem, to relate the forward Compton ${\rm Im\,} H_i$ to inelastic scattering cross sections. 
\item A reason for discarding contributions from some 
$\infty$ contour, if the dispersion relation is to be ``unsubtracted.''
\end{itemize}
The first two requirements are not in question.

\begin{figure}[htbp]
\begin{center}
\includegraphics[height=5cm]{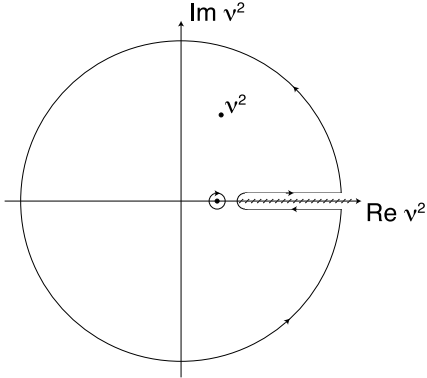}
\caption{Contour in complex $\nu^2$ plane for applying Cauchy identity to $H_1$ or $H_2$.}
\label{fig:cauchy}
\end{center}
\end{figure}

One can consider the elastic and inelastic parts of $H_i$ separately, but it is best to consider them together and separate the terms are the end.  For $H_1$, one can show that it is even in $\nu$, so we will let the dispersion variable be $\nu^2$ rather than $\nu$.    The contour of integration is shown in Fig.~\ref{fig:cauchy},  where one should imagine the outside circle having infinite radius.  Along the real axis, the isolated pole corresponds to elastic scattering and the cut is for inelastic scattering kinematics.  The result for $H_1$ at some general point $\nu^2$ begins its existence as
\ba
H_1(\nu,q^2) 
	= \frac{{\rm Res}  \left. H_1 (\nu,q^2) \right|_{el} }{\nu^2_{\rm el} -\nu^2}
	+ \frac{1}{\pi} \int_{cut} \frac{ {\rm Im\,} H_1(\nu',q^2)}{ {\nu'}^2-\nu^2} d{\nu'}^2
	+ \frac{1}{2\pi i}   \int_{|\nu'|=\infty}  \frac{ H_1(\nu',q^2) }{ {\nu'}^2-\nu^2 } d{\nu'}^2  \,.
\ea

%%%%%%%%%%%%%%%%%%
%%%%%%%%%%%%%%%%%%

The numerator of the first term is the residue (Res) from the poles in $\nu$ for the elastic part of $H_1$, as from Fig.~\ref{boxes}.  

We can interject here that an alternative calculation of the elastic contributions can be done directly, with no dispersion relations, simply using the photon-proton vertex given earlier (Eq.~(\ref{vertex})), whether or not the intermediate proton is on-shell.  We do not recommend doing the calculation this way, since the vertex cannot be guaranteed correct for off-shell protons, but the result is instructive.  For purposes of discussion we quote the result for $H_1$:
\ba
\label{eq:el}
H_1^{\rm el} = - \frac{2M}{\pi} \left( 
	\frac{ q^2 F_1(q^2) G_M(q^2) } { (q^2 + i\epsilon)^2 - 4M^2 \nu^2 }
		+ \frac{ F_2^2(q^2) }{ 4 M^2 }  \right)		\,.
\ea
One can obtain the residue for the dispersion relation from the first term above (and do so correctly, since the result is correct at the pole), and insert it into the dispersion relation to reproduce (of course) the same term.  Of more interest is that the $F_2^2$ term is constant is $\nu$, has no imaginary part, and is therefore absent in the dispersive calculation.  Thus there is a difference, at least at the moment, between the elastic part of the result obtained from the dispersive calculation and what one finds in, for example,~\cite{Bodwin:1987mj} or~\cite{Martynenko:2004bt}.

The second term leads to the $g_1$ term in the quantity $\Delta_1$ given earlier, after using the optical theorem to relate ${\rm Im\,} H_i$ to $g_1$.

The third term is the integral over the part of the contour which is the infinite radius circle.  The commonly quoted results for $\Delta_{\rm pol}$, which appear in this paper, depend on dropping this term. The term is zero, if $H_1$ falls to zero at infinite $|\nu|$.   Assuming this is true, however, appears to be a dramatic assumption.  For example, if the above $H_1^{el}$ were correct (and it could be: the oft criticized vertex in Eq.~(\ref{vertex}) is not guaranteed to be right, but neither are we aware of a guarantee that it is wrong), the assumption would fail for $H_1^{el}$ alone.  Hence, for the assumption to succeed requires an exact cancelation between elastic and inelastic contributions, or a failure of Eq.~(\ref{eq:el}) on the big contour.  On the positive side are several considerations.  One is that nearly the same derivation gives the GDH sum rule, which is checked experimentally and works, within current experimental uncertainty (8\%)~\cite{Pedroni:2006ta}. Also, the GDH sum rule has been checked in lowest order and next-to-lowest order perturbation theory in QED, where it appears to work~\cite{Dicus:2000cd,Altarelli:1972nc}.  Finally, Regge theory suggests the Compton amplitude does fall to zero with energy~\cite{abarb67}, as one would like, although Regge theory famously gave wrong high $\nu$ behavior for spin-independent analogs of $g_1$ and $g_2$~\cite{Damashek:1969xj}.  On the other hand, parton model calculations~\cite{Brodsky:1973hm} have suggested a reason why the Regge theory would fail for the spin-idependent structure functions but still be correct for the the spin-dependent ones.   Hence there are indications, though not decisive proof, supporting the unsubtracted dispersion relation. 

The dispersive derivation finishes by subtracting a term involving $F_2^2$ from the relativistic recoil term, so as to obtain exactly the elastic corrections $\Delta_{\rm el} = \Delta_Z + \Delta_R^p$ that were obtained (say) by Bodwin and Yennie for a calculation of the elastic terms only, using Eq.~(\ref{vertex}) at the photon-proton vertices and no dispersion theory~\cite{Bodwin:1987mj}.  After adding the same term to the polarizability corrections in $\Delta_1$, one obtains the commonly quoted result for $\Delta_1$~\cite{Drell:1966kk,DeRafael:mc,Faustov:yp}.   As noted earlier, this reorganization also allows $\Delta_1$ to be finite in the $m_e \to 0$ limit.  Beyond the historical connection, if one is comfortable with the unsubtracted dispersion relation, the use of the dispersion theory gives a more secure result because it uses only the pole part of the photon-proton-proton vertex, so that the combined elastic and inelastic result does not depend on the general validity of whatever photon-proton-proton vertex one uses.

%%%%%%%%%%%%%%%%%%%%%%%%%%%%%%%%%%%%%%%%%%%%%%%%%%%%%%%%%%%%%%%%%%%%%%%%%%%%

\subsection{Remarks and prospects}

%%%%%%%%%%%%%%%%%%%%%%%%%%%%%%%%%%%%%%%%%%%%%%%%%%%%%%%%%%%%%%%%%%%%%%%%%%%%

Thus the calculated hyperfine splitting in atomic hydrogen is an example of two-photon physics, and requires proton structure information measured at nuclear and particle physics laboratories.  Until 2006, the largest uncertainty was in the proton polarizability corrections, which are related to data from polarized inelastic electron-proton scattering.  The numerical value of the polarizability contributions to hydrogen hyperfine structure, based on latest proton structure function data is
$
\Delta_{\rm pol} = ( 1.3 \pm 0.3 ) {\rm\ ppm}
$.
This is quite similar to the Faustov-Martyenko 2002 result,  
which we think is remarkable given the improvement in the data upon which it is based.  Most of the calculated $\Delta_{\rm pol}$ comes from integration regions where the photon four-momentum squared is small, $Q^2 < 1$ GeV$^2$.

There is still a modest discrepancy between what we think are the best hydrogen hfs calculations and experiment, on the order of 2 ppm.  Where may the problem lie?  It could be in the use of the unsubtracted dispersion relation; or it could be in the value of the  Zemach radius, which taken at face value now contributes the largest single uncertainty among the hadronic corrections to hfs; or perhaps it is a low $Q^2$ surprise in $g_1$ or $g_2$.     It is at any rate not a statistical fluctuation in the hfs data itself.

An interplay between the fields of atomic and nuclear or particle physics may be relevant to sorting out the problem.  For one example, the best values of the proton charge radius currently come from small corrections accurately measured in atomic Lamb shift~\cite{Mohr:2000ie}.  The precision of the atomic measurement of the proton charge radius can increase markedly if the Lamb shift is measured in muonic hydrogen~\cite{Antognini:2005fe}, and data may be taken in 2007 at the Paul Scherrer Institute.  In the present context, the charge radius is noticed by its effect on determinations of the Zemach radius.

We close this section by noting that the final EG1 data analysis from JLab/CLAS should be released soon, and this may shift the value of $\Delta_{\rm pol}$ somewhat.  We may also note that one can keep the lepton masses so as to calculate muonic hydrogen hyperfine splitting, and calculations have already appeared~\cite{Faustov:2006ve,Faustov:2001pn}.

\section{Beam and target normal spin asymmetries}
\label{sec:beam}

In this section we discuss the imaginary part of the two-photon exchange 
amplitude. It can be accessed in the beam or target normal spin asymmetries 
in elastic electron-nucleon scattering,   
and measures the non-forward structure functions of the nucleon.  
After briefly reviewing the theoretical formalism, we discuss  
calculations in the threshold region, in the resonance region, 
in the diffractive region, corresponding with high energy and 
forward angles, as well as in the hard scattering region.  
\newline
\indent
The imaginary (absorptive) part of the $2 \gamma$ exchange amplitude 
can be accessed through a single spin asymmetry (SSA) in 
elastic electron-nucleon scattering, when either the target or beam spin 
are polarized normal to the scattering plane, as has been discussed some time 
ago~\cite{Barut60,RKR71,RKR73}. As time reversal invariance forces 
this SSA to vanish for one-photon exchange, it 
is of order $\alpha = e^2 / (4 \pi) \simeq 1/ 137$. Furthermore, to polarize   
an ultra-relativistic particle in the direction 
normal to its momentum involves 
a suppression factor $m / E$ (with $m$ the mass and $E$ the 
energy of the particle), 
which for the electron is of order $10^{-4} - 10^{-3}$ when the electron 
beam energy is in the 1 GeV range. Therefore, the 
resulting target normal SSA can be expected to be of order $10^{-2}$, 
whereas the beam normal SSA is of order $10^{-6} - 10^{-5}$.
A measurement of such small asymmetries is quite demanding experimentally. 
However, in the case of a polarized lepton beam, asymmetries of the order ppm 
are currently accessible in parity violation (PV) elastic $eN$  
scattering experiments. 
The parity violating asymmetry involves a beam spin polarized 
along its momentum. However the SSA for an electron 
beam spin normal to the scattering plane can also be measured using the 
same experimental set-ups. 
First measurements of this beam normal SSA at beam energies up to 1 GeV 
have yielded values around $-10$ ppm~\cite{Wells01,Maas03,Kaufmann}  
in the forward angular range and up to an order of magnitude 
larger in the backward angular range~\cite{Capozza}. 
At higher beam energies, first results for 
the beam normal SSA in elastic electron-nucleon scattering 
experiments have also been reported recently~\cite{Kaufmann,E158,G0}. 
\newline
\indent
First estimates of the target normal SSA in elastic electron-nucleon 
scattering have been performed in~\cite{RKR71,RKR73}. 
In those works, the $2 \gamma$ exchange 
with nucleon intermediate state (so-called elastic or nucleon 
pole contribution) has been calculated, and the inelastic contribution 
has been estimated in a very forward angle approximation. 
Estimates within this approximation have also been reported for the 
beam normal SSA~\cite{AAM02}. 
The general formalism for elastic electron-nucleon scattering with 
lepton helicity flip, which is needed to describe the beam normal SSA, 
has been developed in~\cite{GGV04}. 
Furthermore, the beam normal SSA has also been estimated at large 
momentum transfers $Q^2$ in~\cite{GGV04} using a parton model, 
which was found crucial \cite{YCC04} to interpret the 
results from unpolarized electron-nucleon elastic scattering, 
as discussed in Section~2.  
In the handbag model of Refs.~\cite{YCC04,ABCCV05,GGV04}, 
the corresponding $2 \gamma$ exchange amplitude 
has been expressed in terms of generalized parton distributions, and 
the real and imaginary part of the 
$2 \gamma$ exchange amplitude are related through a dispersion relation. 
Hence in the partonic regime, 
a direct comparison of the imaginary part with experiment 
can provide a very valuable cross-check 
on the calculated result for the real part.  
\begin{figure}
\begin{center}
\resizebox{0.3\textwidth}{!}{
  \includegraphics{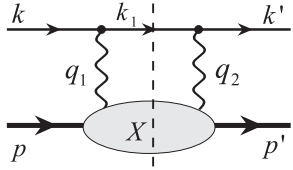}
}
\end{center}
\vspace{-0.5cm}
\caption{The two-photon exchange diagram. 
The filled blob represents the response
of the nucleon to the scattering of the virtual photon.
In the imaginary part of the two-photon amplitude, the intermediate 
state indicated by the vertical dashed line is on-shell.}
\label{fig:2gamma}
\end{figure}
\newline
\indent
To use the elastic electron-nucleon scattering at low momentum 
transfer as a high precision tool, such as 
in present day PV experiments, one may also want to quantify the 
$2 \gamma$ exchange amplitude. To this aim, one may envisage 
a dispersion formalism for the elastic electron-nucleon scattering 
amplitudes, as has been discussed some time ago 
in the literature~\cite{Penarrocha:1980fx,Bordes:1986km}. 
To develop this formalism, the necessary first step 
is a precise knowledge of the imaginary part of the two-photon exchange 
amplitude, which enters in both the beam and target normal SSA. 
Using unitarity, one can relate the imaginary part of the $2 \gamma$  
amplitude to the electro-absorption amplitudes on a nucleon, 
see Fig.~\ref{fig:2gamma}.

\subsection{Single spin asymmetries in elastic electron-nucleon scattering}
\label{sec:beam1}

An observable which is directly proportional to 
the imaginary part of the 
two- (or multi-) photon exchange is given by the elastic scattering of an
unpolarized electron on a proton target polarized {\it normal} to the
scattering plane (or the recoil polarization normal to the
scattering plane, which is exactly the same assuming 
time-reversal invariance). 
For a target polarized perpendicular to the scattering plane, the
corresponding single spin asymmetry, which we refer to as the target normal
spin asymmetry ($A_n$), is defined by~:
\begin{eqnarray}
A_n \,=\, 
\frac{\sigma_\uparrow-\sigma_\downarrow}{\sigma_\uparrow+\sigma_\downarrow}\,,
\label{eq:tasymm}
\end{eqnarray} 
where $\sigma_\uparrow$ ($\sigma_\downarrow$) denotes the cross section 
for an unpolarized beam and for a nucleon spin 
parallel (anti-parallel) to the normal polarization vector, defined
as~:
\begin{eqnarray}
S_n^\mu \,=\, (\,0\,,\, \vec S_n \,), \hspace{1cm}
\vec S_n \,\equiv \,  (\vec{k}\times\vec{k}') \,/\, | \vec{k}\times\vec{k}' | .
\label{eq:sn}
\end{eqnarray}
\indent
As has been shown by de Rujula {\it et al.} \cite{RKR71}, 
the target (or recoil) normal spin asymmetry 
is related to the absorptive part of the elastic $e N$ scattering
amplitude as~:
\begin{equation}
A_n\;=\;
\frac{2 \, {\rm Im}(\sum_{spins}T_{1\gamma}^*\cdot {\rm Abs} \,T_{2\gamma})}
{\sum_{spins}|T_{1\gamma}|^2}\, , 
\label{eq:an1}
\end{equation}
where $T_{1\gamma}$ denotes the one-photon exchange amplitude. 
Since the one-photon exchange amplitude 
is purely real, the leading contribution to $A_n$ is of order
$O(e^2)$, and is due to an interference between one- and two-photon
exchange amplitudes. 
When neglecting terms which correspond with electron helicity flip 
(i.e. setting $m_e = 0$), 
$A_n$ can be expressed in terms of the 
invariants for electron-nucleon elastic scattering, 
defined in Eq.~(\ref{eq:tmatrix}) as~\cite{YCC04}~:
\begin{eqnarray}
A_n &=& \sqrt{\frac{2 \, \varepsilon \, (1+\varepsilon )}{\tau}} \,\,
\left( G_M^2 \,+\, \frac{\varepsilon}{\tau} \, G_E^2 \right)^{-1} \nonumber\\
&\times& \left\{ - \, G_M \, {\cal I} 
\left(\delta \tilde G_E + \frac{\nu}{M^2} \tilde F_3 \right) 
\,+ \, G_E \, {\cal I} \left(\delta \tilde G_M 
+ \left( \frac{2 \varepsilon}{1 + \varepsilon} \right) 
\frac{\nu}{M^2} \tilde F_3 \right) \right\}  
+ {\mathcal{O}}(e^4) , 
\label{eq:tnsa}
\end{eqnarray}
where $\cal I$ denotes the imaginary part. 
\newline
\indent
For a beam polarized perpendicular to the scattering plane, one can
also define a single spin asymmetry, 
analogously as in Eq.~(\ref{eq:tasymm}) as noted in Ref.~\cite{AAM02}, 
where now 
$\sigma_\uparrow$ ($\sigma_\downarrow$) denotes the
cross section for an unpolarized target and for an electron beam spin 
parallel (anti-parallel) to the normal polarization vector, given by 
Eq.~(\ref{eq:sn}). 
We refer to this asymmetry as the beam normal
spin asymmetry ($B_n$). It explicitly vanishes when $m_e = 0$ as it
involves an electron helicity flip. The general electron-nucleon
scattering amplitude including lepton helicity flip involves six invariant
amplitudes and has been worked out in Ref.~\cite{GGV04}, where the 
expression for $B_n$ can also be found. 
As for $A_n$, also $B_n$ vanishes in the Born approximation,
and is therefore of order $e^2$. 

\subsection{Imaginary (absorptive) part of the 
two-photon exchange amplitude}
\label{sec:beam2}

In this section we discuss the relation between the imaginary part of the
two-photon exchange amplitude and the absorptive part 
of the doubly virtual Compton scattering tensor on the nucleon, 
as shown in Fig.~\ref{fig:2gamma}. 
\newline
\indent
The discontinuity of the two-photon exchange amplitude, 
shown in Fig.~\ref{fig:2gamma}, can then be expressed as~:
\begin{equation}
{\rm Abs} T_{2\gamma} = e^4\int\frac{d^3\vec{k}_1}{(2\pi)^32E_{k_1}}
\bar{u}(k',h')\gamma_\mu(\gamma \cdot k_1+m_e)\gamma_\nu u(k,h) 
\cdot
\frac{1}{Q_1^2Q_2^2}\cdot W^{\mu\nu}(p',\lambda_N';p,\lambda_N) \,,
\label{eq:abs}
\end{equation}
\noindent
where the momenta are defined as indicated on Fig.~\ref{fig:2gamma}, 
with $q_1 \equiv k - k_1$, $q_2 \equiv k' - k_1$, and $q_1 - q_2 = q$.
Here $h (h')$ denote the helicities of the initial (final) electrons
and $\lambda_N (\lambda_N')$ denote the helicities of the 
initial (final) nucleons. 
In Eq.~(\ref{eq:abs}), the 
hadronic tensor $W^{\mu\nu}(p',\lambda_N';p,\lambda_N)$ 
corresponds with the absorptive part
of the doubly virtual Compton scattering tensor 
with two {\it space-like} photons~:
\begin{equation}
W^{\mu\nu}(p',\lambda_N';p,\lambda_N) =
\sum_X \,(2\pi)^4 \, \delta^4(p+q_1-p_X) \, 
<p', \lambda_N' |J^{\dagger \mu}(0)|X>  <X|J^\nu(0)|p , \lambda_N>,
\label{eq:wtensor} 
\end{equation}
\noindent
where the sum goes over all possible {\it on-shell} intermediate hadronic 
states $X$ (denoting $p_X^2 \equiv W^2$). 
Note that in the limit $p' = p$, Eq.~(\ref{eq:wtensor}) 
reduces to the forward tensor for inclusive electron-nucleon scattering 
and can be parametrized by the usual 4 nucleon forward structure 
functions. In the non-forward case however, 
the absorptive part of the 
doubly virtual Compton scattering tensor of Eq.~(\ref{eq:wtensor}) 
which enters in the evaluation of target and beam normal spin asymmetries, 
depends upon 18 invariant amplitudes~\cite{Tar75}. 
Though this may seem as a forbiddingly large number of new functions, 
we may use the unitarity relation to express the full non-forward tensor 
in terms of electroproduction amplitudes $\gamma^* N \to X$. 
The number of intermediate states $X$ which one considers in the 
calculation will then put a limit on how high in energy one can 
reliably calculate the hadronic tensor Eq.~(\ref{eq:wtensor}). 
In the following section, the tensor $W^{\mu\nu}$ will be discussed for the
elastic contribution ($X = N$), in the resonance region as 
a sum over all $\pi N$ intermediate states (i.e. $X = \pi N$), 
using a phenomenological state-of-the-art calculation for the 
$\gamma^* N \to \pi N$ amplitudes, in the 
diffractive region (corresponding with high energy, forward scattering) 
where it can be related to the 
total photo-absorption cross section on a proton, as well as in the 
hard scattering region where it can be related to nucleon generalized 
parton distributions.   
\newline
\indent
There are special regions in the phase space integral 
of Eq.~(\ref{eq:abs}), corresponding 
with near singularities, which may give important contributions 
(logarithmic enhancements) under some kinematical conditions. 
When the intermediate and initial electrons are collinear, 
then also the photon with momentum $\vec q_1$ is 
collinear with this direction. 
For the elastic case ($W = M$) this precisely corresponds with 
the situation where the first photon is soft (i.e. $q_1 \to 0$) and 
where the second photon carries the full momentum transfer 
$Q_2^2 \simeq Q^2$.  
For the inelastic case ($W > M$) the first photon is 
hard but becomes quasi-real (i.e. $Q_1^2 \sim m_e^2$). 
In this case, the virtuality of the second photon is smaller than $Q^2$. 
An analogous situation occurs when the intermediate electron is 
collinear with the final electron. 
These kinematical situations with one quasi-real photon and one virtual photon 
correspond with quasi virtual Compton scattering (quasi-VCS).  
Besides the quasi-VCS singularities, the two-photon exchange 
amplitude also has a near singularity when 
the intermediate electron momentum is soft (i.e. $|\vec k_1| \to 0$).  
In this case both photons are hard but have virtualities which become very 
small, and vanish if the electron mass is taken to zero. 
This situation, with two quasi-real photons, 
occurs when the invariant mass of the hadronic 
state takes on its maximal value $W_{max} = \sqrt{s} - m_e$, and    
corresponds to quasi-real Compton scattering (quasi-RCS).

\subsection{Results and discussion}
\label{sec:beam3}

\subsubsection{Threshold region}

\begin{figure}[h]
\leftline{\resizebox{0.4\textwidth}{!}{
\includegraphics{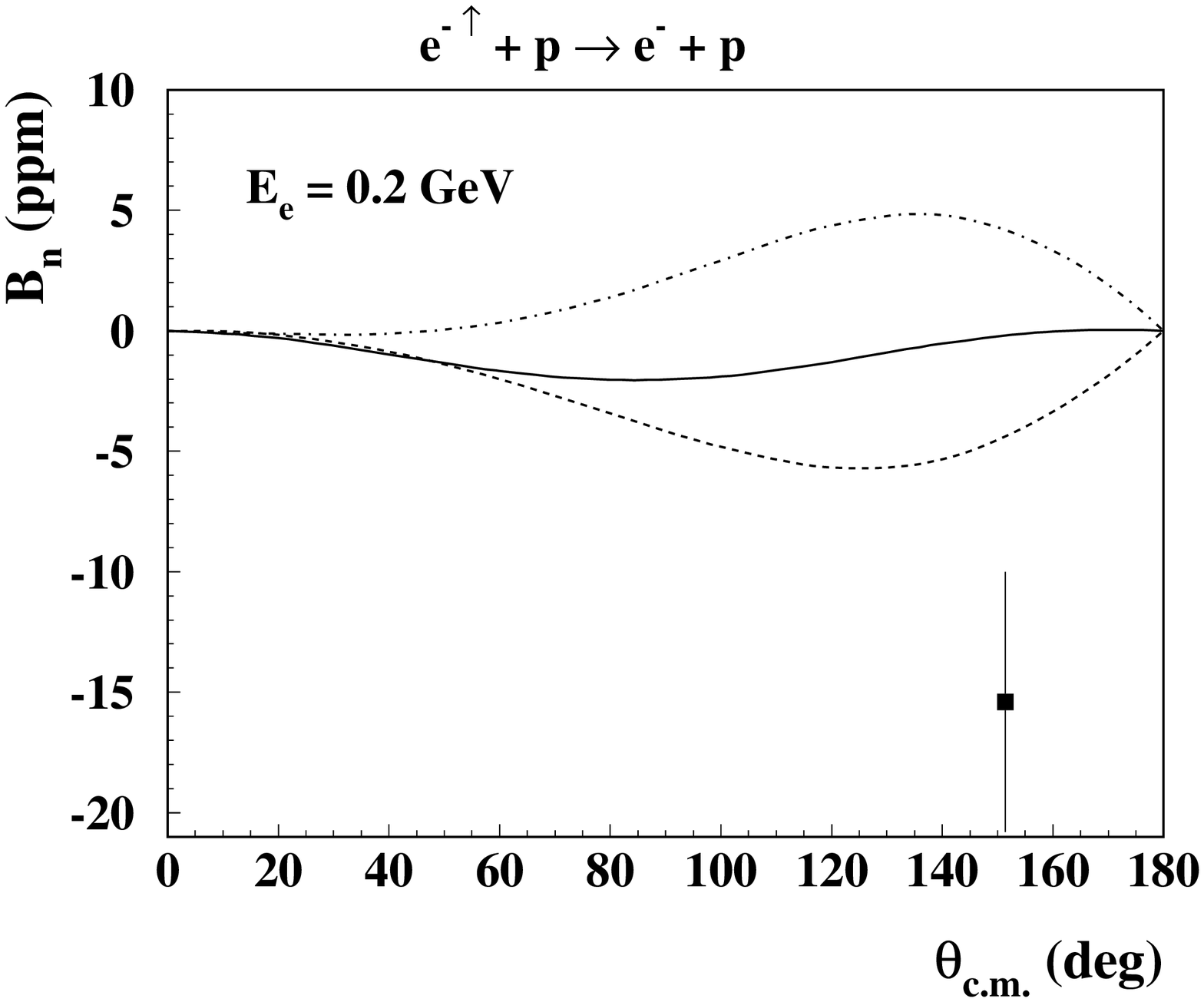}
}}
\vspace{-5.5cm}
\rightline{\resizebox{0.45\textwidth}{!}{
\includegraphics{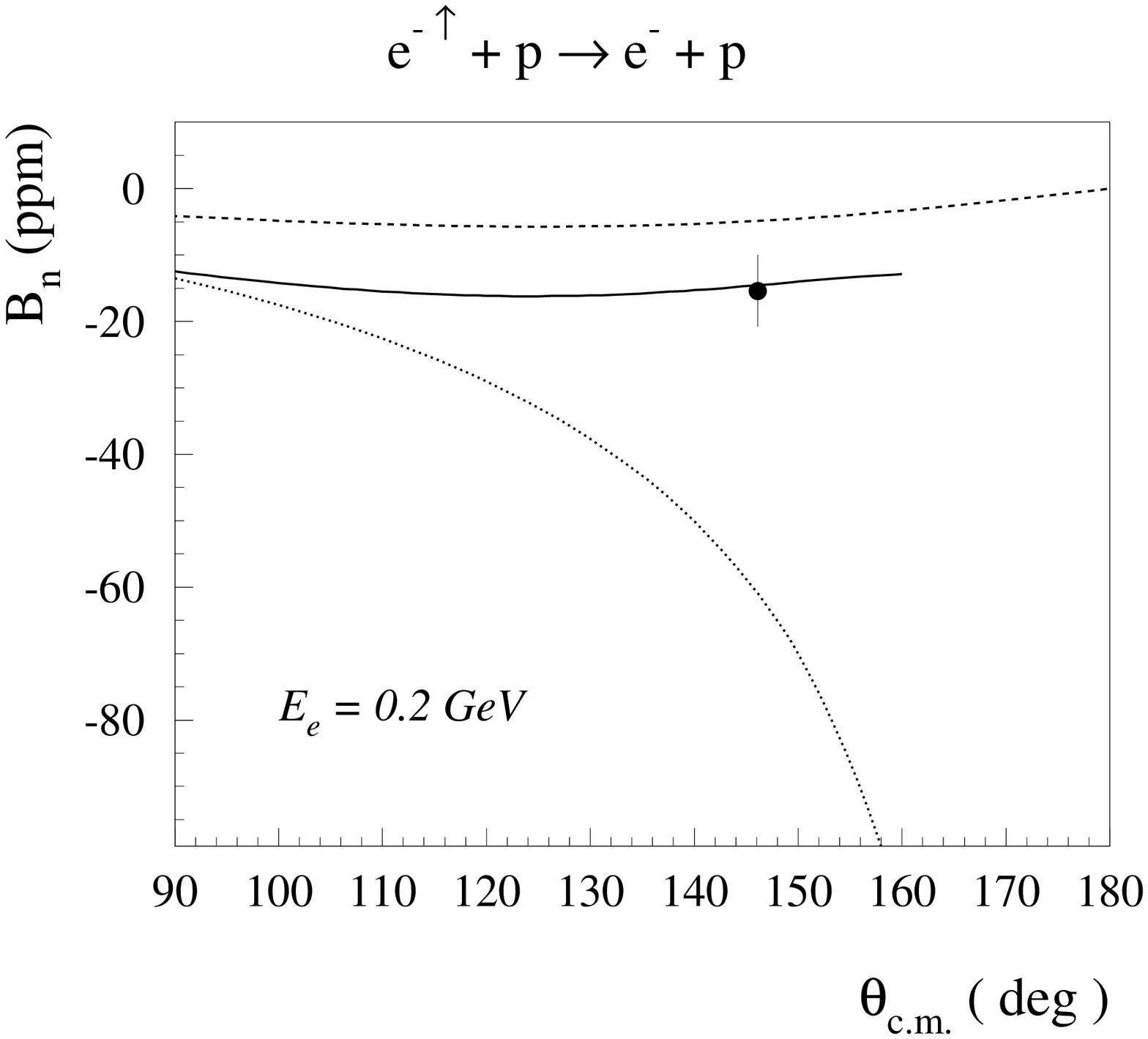}
}}
\caption{Beam normal spin asymmetry for $e^{- \uparrow} p \to e^- p$ 
at a beam energy $E_e = 0.2$~GeV 
as function of the $c.m.$ scattering angle. Left panel :   
calculation from Ref.~\cite{Pasquini:2004pv} 
for different hadronic intermediate states ($X$) in the blob  of
Fig.~\ref{fig:2gamma} ~:  $N$ (dashed curve),  
$\pi N$  (dashed-dotted curve), 
sum of the $N$ and $\pi N$ (solid curve). 
Right panel :  calculation of Ref.~\cite{Diaconescu:2004aa}, where the 
nucleon intermediate state is expanded to 
leading order (dotted curve) and next to leading order (solid curve) 
in $E_e / M$. For comparison, also the full nucleon intermediate state
result (dashed curve, same as on left panel) is shown. 
The data point is from the SAMPLE Coll.~\cite{Wells01}.}
\label{fig:23}     
\end{figure}

In Ref.~\cite{Diaconescu:2004aa}, the beam normal spin asymmetry 
was studied at low energies in an effective theory of electrons, protons and 
photons. This calculation, in which pions are integrated out, effectively 
corresponds with the nucleon intermediate state contribution only, 
expanded to second order in $E_e / M$. To this order, 
the calculation includes the recoil corrections to the scattering from 
a point charge, the nucleon charge radius, and the nucleon isovector magnetic 
moment. One sees from Fig.~\ref{fig:23} (right panel) 
that the theory expanded up to second 
order in $E_e / M$ (indicated by the full results) is able to give a good 
account of the SAMPLE data point at the low energy $E_e = 0.2$~GeV. 
\newline
\indent
However, when doing the full calculation for the 
$N$ intermediate state, which is 
model independent (as it only involves on-shell $\gamma^* NN$ matrix 
elements), the result is further reduced as seen in 
Fig.~\ref{fig:23} (left panel). 
Inclusion of threshold pion electroproduction contributions, arising from the 
$\pi N$ intermediate states, partly cancels the elastic contributions. 
Because in this low-energy region, the matrix elements are 
rather well known, it is not clear at present how to get a better agreement 
with the rather large asymmetry measured by SAMPLE~\cite{Wells01}.

\subsubsection{Resonance region}

When measuring the imaginary part of the elastic $eN$ 
amplitude through a normal SSA at sufficiently low energies,  
below or around two-pion production threshold, 
one is in a regime where these electroproduction amplitudes are relatively 
well known using pion electroproduction experiments as input. 
As both photons in the $2 \gamma$ exchange process are 
virtual and integrated over, an observable such as the beam or target normal 
SSA is sensitive to the electroproduction amplitudes 
on the nucleon for a range of photon virtualities. 
This may provide information on resonance transition 
form factors complementary to the information obtained from current 
pion electroproduction experiments.  
\newline
\indent
In Ref.~\cite{Pasquini:2004pv}, the imaginary part of the two-photon exchange 
amplitude was calculated by relating it through unitarity to the contribution 
of $X = N$ and $X = \pi N$ intermediate state contributions. For the $\pi N$ 
intermediate state contribution, the corresponding pion electroproduction 
amplitudes were taken from the phenomenological MAID analysis~\cite{Dre99}, 
which contains both resonant and non-resonant pion production mechanisms. 
The calculation of \cite{Pasquini:2004pv} shows that 
at forward angles, the quasi-real 
Compton scattering at the endpoint $W = W_{max}$ only yields a very 
small contribution, which grows larger when going to backward angles. 
This quasi-RCS contribution is of opposite sign as the 
remainder of the integrand, and therefore determines the position of the 
maximum (absolute) value of $B_n$ when going to backward angles. 
\begin{figure}[h]
\vspace{-1.5cm}
\begin{center}
\resizebox{0.55\textwidth}{!}{
  \includegraphics{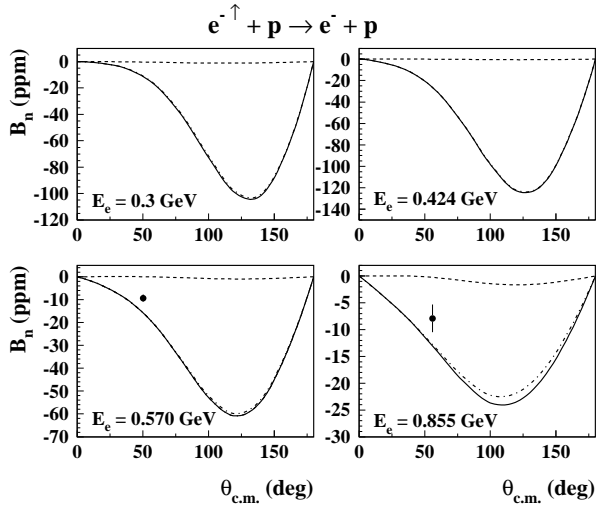}
}
\end{center}
\vspace{-1cm}
\caption{$B_n$ for $e^{- \uparrow} p \to e^- p$ 
as function of the $c.m.$ scattering angle  
at different beam energies, as indicated on the figure.  
The calculations are for different hadronic intermediate 
states ($X$) in the blob of Fig.~\ref{fig:2gamma} ~: 
$N$ (dashed curve),  
$\pi N$  (dashed-dotted curves), 
sum of the $N$ and $\pi N$ (solid curves). 
The data points are from the 
A4 Coll. (MAMI)~\cite{Maas03}. 
Calculations and figure from Ref.~\cite{Pasquini:2004pv}.}
\label{fig:5}
\end{figure}
\newline
\indent
In Fig.~\ref{fig:5}, the results for $B_n$ are shown 
at different beam energies below $E_e = 1$~GeV. 
It is clearly seen that at energies $E_e = 0.3$~GeV and higher  
the nucleon intermediate state (elastic part) 
yields only a very small relative 
contribution. Therefore $B_n$ is a direct measure of the inelastic part which 
gives rise to sizeable large asymmetries, of the order of several tens of ppm 
in the backward angular range, mainly driven by the quasi-RCS 
near singularity. First results from the A4 Coll. for $B_n$ 
at backward angles (for $E_e$ around 0.3~GeV)  
indeed point towards a large $B_n$ value of order -100 ppm 
for $\theta_{cm}$ around 150~deg~\cite{Capozza}.  
At forward angles, the sizes of the predicted  
asymmetries are compatible with the first 
high precision measurements performed by the A4 Coll.~\cite{Maas03}, 
though the model slightly overpredicts (in absolute value) $B_n$ at  
$E_e = 0.570$~GeV and 0.855~GeV.

\subsubsection{High-energy, forward scattering (diffractive) region}

\begin{figure}[h]
\vspace{-1cm}
\begin{center}
\resizebox{0.8\textwidth}{!}{
\includegraphics{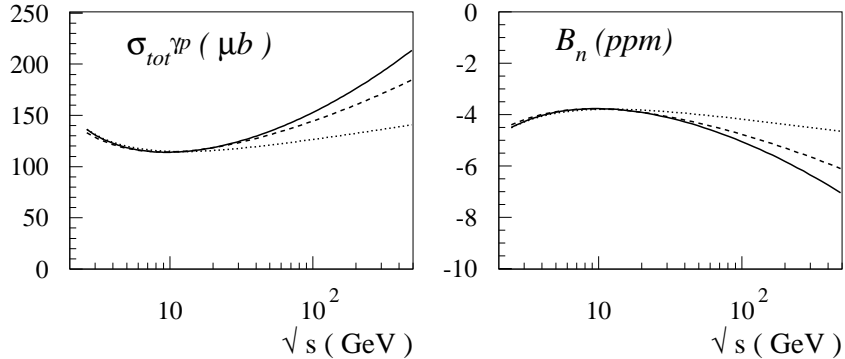}
}
\end{center}
\vspace{-1.5cm}
\caption{Energy dependence of 
$B_n$ for $e^{- \uparrow} p \to e^- p$ (right panel) 
at $Q^2 = 0.05~$GeV$^2$ (corresponding with very forward scattering angle) 
using Eq.~(\ref{eq:diffr}) for different parameterizations of the 
total photo-absorption cross section on the proton (left panel)~:
Block-Halzen $log$ fit (dotted curves), 
Block-Halzen $log^2$ fit (solid curves), and 
Donnachie-Landshoff fit (dashed curves).   
Calculations from Ref.~\cite{Afanasev:2004pu}.}
\label{fig:6}  
\end{figure}

At very high energies and forward scattering angles (so-called diffractive 
limit), it was shown in 
Refs.~\cite{Afanasev:2004pu,Gorchtein:2005za} that $B_n$ 
is dominated by the quasi-real Compton singularity. 
In this (extreme forward limit) case, the hadronic tensor 
can be expressed in terms of the total photo-absorption cross section 
on the proton, $\sigma_{tot}^{\gamma p}$,  
allowing to express $B_n$ through the simple analytic expression~:
\begin{eqnarray}
B_n = - \frac{m_e \, \sqrt{Q^2} \, \sigma_{tot}^{\gamma p}}{8 \pi^2} \,
\frac{G_E}{\tau G_M^2 + \varepsilon G_E^2} \, 
\left[ \log \frac{Q^2}{m_e^2} - 2 \right].
\label{eq:diffr}
\end{eqnarray}
One notices that the quasi-real Compton singularity gives rise to a (single) 
logarithmic enhancement factor which is at the orgin of the relatively large 
value of $B_n$. 
\newline
\indent
In Fig.~\ref{fig:6}, the estimate from Ref.~\cite{Afanasev:2004pu} 
based on Eq.~(\ref{eq:diffr}) is shown 
for different parameterizations of the total photo-absorption cross section. 
The beam normal spin asymmetry has been measured at SLAC (E-158) at an 
energy $E_e = 46$~GeV ($\sqrt{s} \simeq 9$~GeV) and very forward angle 
($Q^2 \simeq 0.05$GeV$^2$). First result~\cite{E158}     
indicate a value $B_n \simeq -3.5 \to -2.5$~ppm, confirming the estimate shown 
in Fig.~\ref{fig:6}. 
\newline
\indent
At intermediate energies, around $E_e \simeq 3$~GeV, and forward angles, 
$B_n$ has also been measured by the HAPPEX and G0 Collaborations. 
The simple ``diffractive'' formula of Eq.~(\ref{eq:diffr}) does not rigorously 
apply any more and one has to calculate corrections due to the deviation from 
forward scattering. Such calculation have recently been performed in 
Refs.~\cite{Afanasev:2004pu,Gorchtein:2006mq} in different model approaches,
where the calculation of Ref.~\cite{Gorchtein:2006mq} includes subleading 
terms in $Q^2$.    
The predicted asymmetries are in basic agreement with first results 
reported by HAPPEX~\cite{Kaufmann} and G0~\cite{G0}. 
\newline
\indent
In Table~\ref{tab:1}, the present status of measurements of  
beam normal spin asymmetries, most of them in the forward angular range, 
is shown. 
\begin{table}
\caption{Summary of measurements of the beam normal spin asymmetry 
in elastic electron-proton scattering}
\label{tab:1}      
\begin{center}
\begin{tabular}{l|l|l|l}
\hline\noalign{\smallskip}
EXP. & $E_e$ (GeV) & $Q^2$ (GeV$^2$) & $B_n$ (ppm) \\
\noalign{\smallskip}\hline\noalign{\smallskip}
SAMPLE~\cite{Wells01} & 0.192 & 0.10 & -16.4 $\pm$ 5.9 \\
A4~\cite{Maas03} & 0.570 & 0.11 & -8.59 $\pm$ 0.89  \\
A4~\cite{Maas03} & 0.855 & 0.23 & -8.52 $\pm$ 2.31  \\
HAPPEX~\cite{Kaufmann} & 3.0 & 0.11 & -6.7 $\pm$ 1.5\\
G0~\cite{G0} & 3.0 & 0.15 & -4.04 $\pm$ 1.05 \\
G0~\cite{G0} & 3.0 & 0.25 & -4.81 $\pm$ 2.03 \\
E-158~\cite{E158} & 46 & 0.06 & -3.5 $\to$ -2.5 \\
\noalign{\smallskip}\hline
\end{tabular}
\end{center}
\end{table}

\subsubsection{Hard scattering region}

\begin{figure}[h]
\begin{center}
\includegraphics[height=5.5cm]{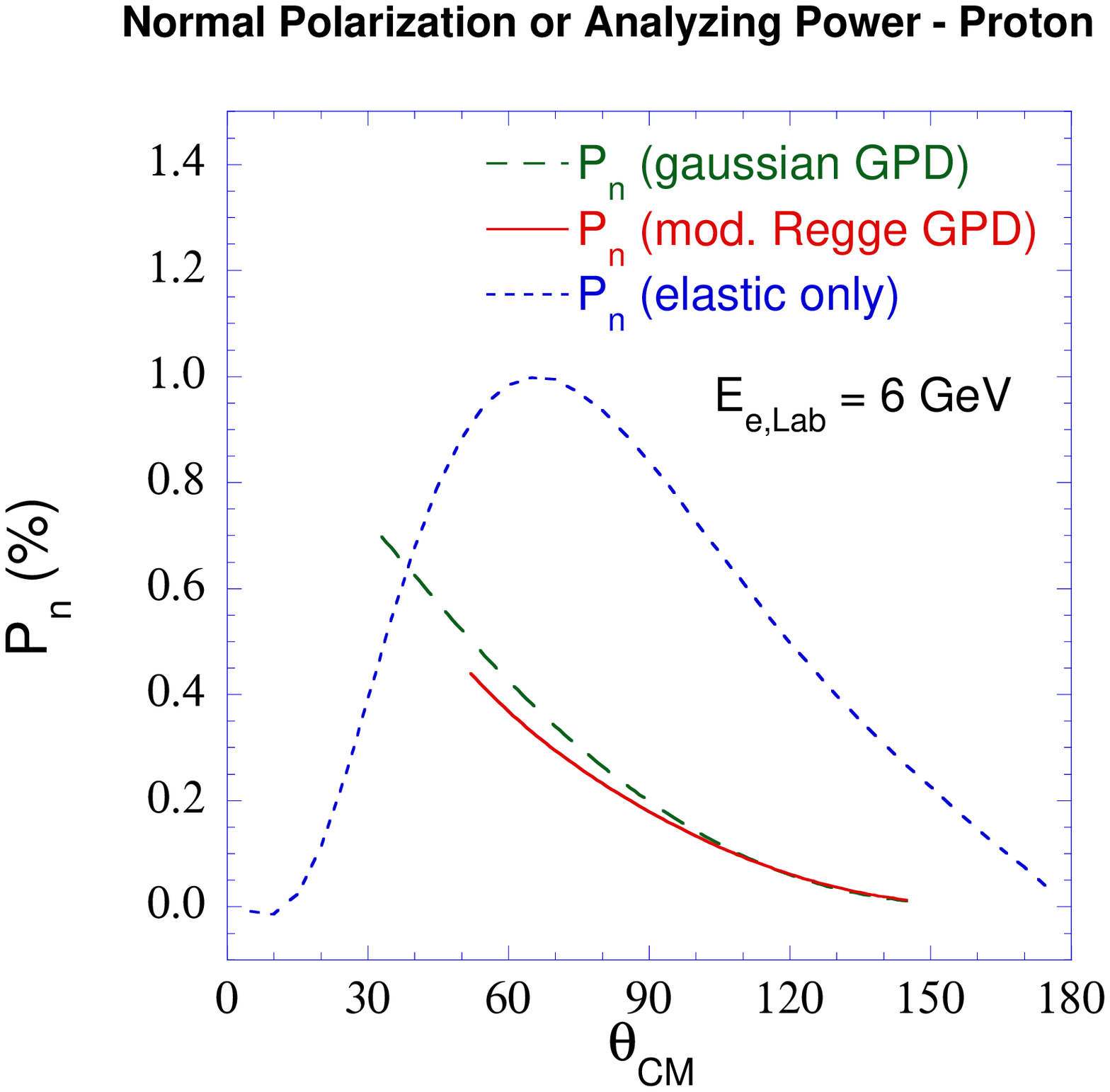} 
\includegraphics[height=5.5cm]{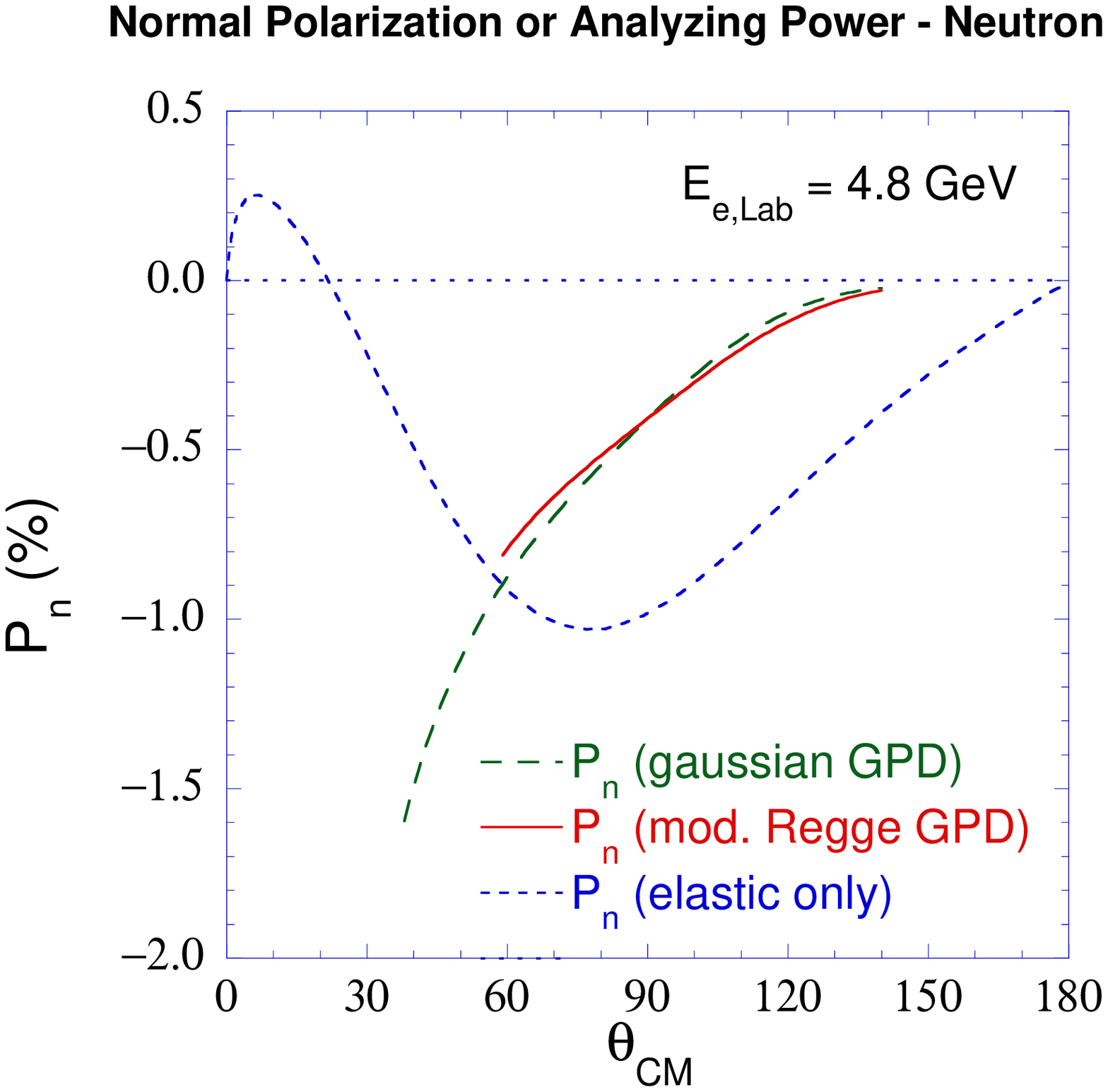}
\end{center}
\vspace{-0.75cm}
\caption{Nucleon analyzing power $A_n$, 
which is equal to the normal polarization $P_n$.
The elastic contribution is shown by the dotted curve~\cite{RKR71}.
The GPD calculation for the inelastic contribution is 
shown by the dashed curve for a gaussian GPD model, and by the solid curve 
for a modified Regge GPD parameterization.  
The GPD calculation is cut off in the backward direction at $-u = M^2$.  
In the forward direction it goes down to 
$Q^2 = 2$ GeV$^2$ (modified Regge GPD) and to $Q^2 = M^2$ (Gaussian GPD). 
Figure from~\cite{ABCCV05}. 
}
\label{fig:7}
\end{figure}

In the hard scattering region, the beam and target normal spin asymmetries  
were estimated in Refs.~\cite{GGV04,ABCCV05} 
through the scattering off a parton, 
which is embedded in the nucleon through a GPD.  
\newline
\indent
The GPD estimate for the target spin asymmetry $A_n$ for the proton 
is shown in the left-hand plot of Fig.~\ref{fig:7} 
as a function of the CM scattering angle for 
fixed incoming electron lab energy, taken here  as $6$ GeV.  
Also shown is a calculation of $A_n$ including the elastic 
intermediate state only~\cite{RKR71}.  
The result, which is nearly the same for either of the two GPD
parameterizations which were used in Ref.~\cite{ABCCV05}, is of order 1\%.
Fig.~\ref{fig:7} on the right also shows a similar plot of 
$A_n$ for a neutron target.  
The predicted asymmetry is of opposite sign, 
reflecting that the numerically largest term in Eq.~(\ref{eq:tnsa})  
is the one proportional to
$G_M$. The results are again of order 1\% in magnitude, 
though somewhat larger for the neutron than for the proton. 
A precision measurement of $A_n$ is planned at JLab \cite{AnTodd} 
on a polarized $^3He$ target; it will provide access to the elastic 
electron-neutron single-spin asymmetry from two-photon exchange. 

\begin{figure}[h]
\begin{center}
\resizebox{0.41\textwidth}{!}{
  \includegraphics{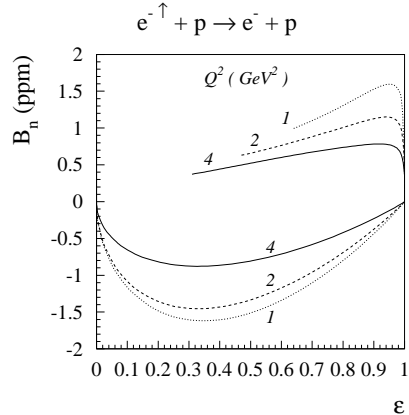}
}
\end{center}
\vspace{-1cm}
\caption{$B_n$ for elastic 
$e^- p$ scattering as function of $\varepsilon$  
at different values of $Q^2$ as indicated on the figure.  
The upper thick curves ($B_n > 0$) 
are the GPD calculations for the kinematical range where $s, -u > M^2$.
For comparison, the elastic contribution is also displayed : lower 
thin curves ($B_n < 0$). Calculations and figure from 
Ref.~\cite{GGV04}.}
\label{fig:8}
\end{figure}

Using the same phenomenological parametrizations for the GPDs,   
$B_n$ was found to yield values around +1 ppm to +1.5 ppm 
in the few GeV beam energy range, see Fig.~\ref{fig:8}.  
In particular, the forward angular  
range for $e^{- \uparrow} p \to e^- p$ scattering was found to be a  
favorable region to get information on the inelastic part of $B_n$. 
Because in the handbag calculation, real and imaginary 
parts are linked, a direct measurement of $B_n$ 
may yield a valuable cross-check for the 
real part, which was found crucial in understanding the unpolarized 
cross section data for $e^- p \to e^- p$ at large momentum transfer.  

\section{Conclusions and open issues}
\label{conclude}

The striking difference between the unpolarized (Rosenbluth) and 
polarization transfer measurements of the proton $G_E / G_M$ form 
factor ratio has triggered a renewed interest in the field of 
two-photon exchange in electron-nucleon scattering experiments. 
Theoretical calculations both within a hadronic and partonic framework, 
which were reviewed in this work,  
made it very likely that hard two-photon exchange corrections 
are the main culprit in the difference between both experimental techniques. 
Despite the long history of two-photon exchange corrections, 
it is interesting to note that concepts developed over the past decade, such
as generalized parton distributions which describe two-photon processes 
with one or two large photon virtualities, enter when quantifying 
two-photon exchange corrections at larger $Q^2$. 

The model-independent finding is that the hard two-photon corrections  
hardly affect polarization transfer results, but they 
do correct the slope of the Rosenbluth plots at larger $Q^2$ in an important
way, towards reconciling the Rosenbluth data with the 
$G_E / G_M$ ratio, which decreases linearly in $Q^2$, 
from the polarization data. 
This being said, it is fair to state at this point that neither 
of the current calculations convincingly quantifies the effect 
either due to uncontrolled approximations 
(such as using ad hoc assumptions for off-shell vertices in hadronic
calculations),  
or extrapolations beyond their region of validity (such as applying partonic
calculations in the low momentum transfer regime). 
The present and forthcoming high-precision electron scattering data, 
which aim  
at testing two-photon exchange effects, have shifted
the emphasis, however, from the qualitative to the quantitative realm. 

Besides entering as corrections to elastic electron-nucleon scattering 
data, two-photon exchange processes are also the leading corrections 
to improve our quantitative understanding of the hyperfine splitting (hfs) in
hydrogen. We reviewed in this work that our best estimate, 
based on the recent data for polarized nucleon structure 
functions which enter the polarizability correction to the hydrogen hfs, 
yields a result 
$\Delta_{pol} = (1.3 \pm 0.3)$~ppm. Combined with recent estimates of the 
Zemach radius, the correction to the hydrogen hfs falls short
of the data by about 2~ppm, or about 2.8 standard deviations. 

We have also reviewed the recent measurements of large beam normal single 
spin asymmetries (SSA), 
of order $-10$ to $-100$~ppm, and measured with ppm level accuracy, which 
arose over the past few years 
as an interesting spin-off of the high precision parity violation 
experiments in electron-nucleon scattering. 
We discussed how such asymmetries, which measure the imaginary part 
of the two-photon exchange amplitude, can be expressed through 
non-forward nucleon structure functions and provide a new tool to access
hadron structure information. 
\\
\\
\indent
We end this review by spelling out a few open issues and challenges 
(both theoretical and experimental) in this field~:

\begin{enumerate}

\item {\it Experimental measurements of the two-photon exchange processes} 

In order to use electron scattering as a precision tool, it is clearly 
worthwhile to arrive at a quantitative understanding of two-photon exchange 
processes. This calls for detailed experimental studies, and 
several new experiments are already planned.  
A first type of experiments is to perform high precision unpolarized 
experiments and look for nonlinearities in the Rosenbluth plots. 
A recent study~\cite{Tvaskis:2005ex} found that the 
non-linearities in $\varepsilon$ in current Rosenbluth data stay small 
over a fairly large $Q^2$ range.       
Such experiments however cannot separate that part 
of the two-photon corrections 
which is linear in $\varepsilon$ and which appears 
to be the dominant source of corrections in theoretical calculations.  
The difference between elastic $e^-$ and $e^+$ scattering on a proton target 
directly accesses the real part of the two-photon exchange amplitude 
and the part linear in $\varepsilon$. At
intermediate $Q^2$ values some forthcoming $e^+/e^-$ experiments 
at JLab~\cite{brooks} and VEPP-3~\cite{arrington2}   
will allow one to systematically study and 
quantify two-photon exchange effects. 
Also forthcoming is an experimental check of the predicted small 
effect of two-photon processes on the polarization data 
by measuring the $\varepsilon$ dependence in polarization transfer 
experiments~\cite{charles}. 

\item
{\it Further refinements on the theoretical side : real part of the two-photon
exchange amplitude} 

The calculations of the real part of the two-photon amplitude 
for elastic $e N$ scattering described in this work are performed
either within a hadronic framework at lower $Q^2$ values 
or partonic picture at larger $Q^2$ values. 
The hadronic calculation assumes some off-shell vertex functions 
and performs a four-dimensional integral of the box diagram. 
It would be desirable to have a dispersion relation framework where 
the real part is explicitely calculated as an integral over the 
imaginary part, which can be related to observables 
({\it i.e.} on-shell amplitudes) 
using unitarity. In the forward limit such dispersion relations are
underlying, {\it e.g.}, the evaluation of the hyperfine splitting in hydrogen. 
In the non-forward case, the convergence of such dispersion relations 
would need further study. Within the partonic framework, a full calculation of
the processes with two-active quarks 
(the so-called cat's ears diagrams) remains to be performed.

\item
{\it Systematic exploration of the imaginary part of the
  two-photon exchange amplitude} 

The ongoing experiments for both the beam and target normal SSA 
in elastic $e N$ scattering will trigger further theoretical work, 
and a cross-fertilization between theory and 
experiment can be expected in this field.

\item
{\it Two-photon exchange effects in resonance production processes}

The sensitivity of unpolarized measurements of $G_E/G_M$ to two-photon 
exchange effects also signals that such potential effects can be expected when 
measuring resonance transition form factors out to larger $Q^2$ values as is 
performed, {\it e.g.}, at JLab. In particular    
a first theoretical study of the two-photon exchange contribution to the 
$e N \to e \Delta(1232) \to e \pi N$ process 
with the aim of a precision study 
of the ratios of electric quadrupole (E2) and Coulomb quadrupole (C2) to 
the magnetic dipole (M1) $\gamma^* N \Delta$ transitions, has been 
recently performed~\cite{Pascalutsa:2005es}.  
The two-photon exchange amplitude has been related to the $N \to \Delta$ 
GPDs. It was found that the   
C2/M1 ratio at larger $Q^2$ values 
depends strongly on whether this quantity is obtained
from an interference cross section or from the Rosenbluth-type cross sections, 
in similarity with the elastic, $e N \to e N$, process.   
It will be interesting to confront these results with 
upcoming new Rosenbluth separation data at intermediate $Q^2$ values 
in order to arrive at a precision extraction of the large $Q^2$  
behavior of the $R_{EM}$ and $R_{SM}$ ratios. 

\item
{\it Two-photon exchange effects in deep-inelastic scattering processes}

Two-photon exchange effects were also studied in the past 
in  deep-inelastic scattering processes; see, {\it e.g.},  
Refs.~\cite{Bartels:1973pf,Kingsley:1972fd,Bodwin:1975ty}.  
They can be expected to affect the extraction
of the longitudinal structure function from $L/T$ Rosenbluth type experiments 
in the deep-inelastic scattering region, which remains to be quantified.  

\item
{\it Two-photon exchange effects in parity violating elastic electron-nucleon
  scattering and related study of $\gamma Z$ box diagram effects}

In recent years an unprecedented precision has been achieved in parity
violating electron scattering experiments.  
Two-photon exchange processes may be 
relevant in the interpretation of the current generation of high 
precision parity-violation experiments, in particular for using those data to
determine the strange-quark content of the proton; for a first study see 
Ref.~\cite{Afanasev:2005ex}. 

A related issue in the interpretation of parity violating electron 
scattering experiments 
are the $\gamma Z$ box diagram contributions. Such calculations have 
been performed for atomic parity 
violation~\cite{Marciano:1982mm} corresponding with zero 
momentum transfer as well as in the deep-inelastic scattering 
region~\cite{Bohm:1986na}, 
by calculation the $\gamma Z$ exchange between and electron and a quark. 
A full calculation in the small and intermediate $Q^2$ regime, where many 
current parity violation experiments are performed is definitely 
a worthwhile topic for further research.  

\item
{\it Two-photon exchange effects in hydrogen hyperfine splitting}

More accurate data, and data at lower $Q^2$, is forthcoming on  polarized
inelastic $e$-$p$ scattering.  These data will allow a more  precise
evaluation of the polarizability corrections to hydrogen  hfs.  In addition, a
measurement of hfs in muonic hydrogen may be
possible~\cite{Antognini:2005fe}.  The calculated corrections for  muonic
hydrogen have different weightings because of the muon mass,  and with
calculations such 
as~\cite{Nazaryan:2005zc,Faustov:2006ve,Faustov:2001pn} 
updated with new scattering data, the result  of the measurement 
could be presented as an independent accurate  determination of the 
proton's Zemach radius. 

\end{enumerate}

\section*{\center{Acknowledgements}}

We like to thank 
A. Afanasev, 
S. Brodsky, 
Y.C. Chen, 
M. Gorchtein, 
K. Griffioen, 
P. Guichon, 
V. Nazaryan, 
V. Pascalutsa, 
and 
B. Pasquini 
for several collaborations on the subjects reviewed in this work. 
Furthermore we would like to thank 
A. Afanasev, 
D. Armstrong, 
T. Averett, 
L. Capozza, 
M. Gorchtein, 
F. Maas, 
Ch. Perdrisat, 
and 
M. Ramsey-Musolf
for discussions and correspondence during the course of this work. 
\newline
\indent
The work of C.~E.~C. is supported by the National Science Foundation
under grant PHY-0555600. 
The work of M.~V. is supported in part by DOE grant
DE-FG02-04ER41302 and contract DE-AC05-06OR23177 under
which Jefferson Science Associates operates the Jefferson Laboratory.


\begin{thebibliography}{999}


%%%%%%%%%%%%%%%%%%%%%%%%%%%
%
%      For introduction
%
%%%%%%%%%%%%%%%%%%%%%%%%%%%

\bibitem{Hyde-Wright:2004gh}
  C.~E.~Hyde-Wright and K.~de Jager,
  %``Electromagnetic Form Factors of the Nucleon and Compton Scattering,''
  Ann.\ Rev.\ Nucl.\ Part.\ Sci.\  {\bf 54}, 217 (2004). 
  %[arXiv:nucl-ex/0507001].
  %%CITATION = NUCL-EX 0507001;%%

\bibitem{Arrington:2006zm}
  J.~Arrington, C.~D.~Roberts and J.~M.~Zanotti,
  %``Nucleon electromagnetic form factors,''
  arXiv:nucl-th/0611050.
  %%CITATION = NUCL-TH 0611050;%%

\bibitem{Perdrisat:2006hj}
  C.~F.~Perdrisat, V.~Punjabi and M.~Vanderhaeghen,
  %``Nucleon electromagnetic form factors,''
  arXiv:hep-ph/0612014.
  %%CITATION = HEP-PH 0612014;%%


\bibitem{Arn81}
A.I. Akhiezer, L.N. Rosentsweig, I.M. Shmushkevich,
Sov. Phys. JETP {\bf 6}, 588 (1958);
J. Scofield, Phys. Rev. {\bf 113}, 1599 (1959); 
			{\it ibid.} {\bf 141}, 1352 (1966);
N.~Dombey,
%``Scattering Of Polarized Leptons At High Energy,''
Rev.\ Mod.\ Phys.\  {\bf 41}, 236 (1969);
%%CITATION = RMPHA,41,236;%%
A. I. Akhiezer and M. P. Rekalo, Sov. J. Part. Nucl. {\bf 4}, 277 (1974);
R.~G.~Arnold, C.~E.~Carlson and F.~Gross,
%``Polarization Transfer In Elastic Electron Scattering From Nucleons And
%Deuterons,''
Phys.\ Rev.\ C {\bf 23}, 363 (1981).
%%CITATION = PHRVA,C23,363;%%


\bibitem{Jones00}
M.~K.~Jones {\it et al.}  [Jefferson Lab Hall A Coll.],
%``G(E(p))/G(M(p)) ratio by polarization transfer in  e(pol.) p $\to$ e
%p(pol.),''
Phys.\ Rev.\ Lett.\  {\bf 84}, 1398 (2000). 
%[arXiv:nucl-ex/9910005].
%%CITATION = NUCL-EX 9910005;%%


\bibitem{Gayou02}
O.~Gayou {\it et al.}  [Jefferson Lab Hall A Coll.],
%``Measurement of G(E(p))/G(M(p)) in e(pol.) p $\to$ e p(pol.) to Q**2 =
%5.6-GeV**2,''
Phys.\ Rev.\ Lett.\  {\bf 88}, 092301 (2002).
%[arXiv:nucl-ex/0111010].
%%CITATION = NUCL-EX 0111010;%%


\bibitem{Punjabi:2005wq}
V.~Punjabi {\it et al.},
  Phys.\ Rev.\ C {\bf 71}, 055202 (2005)
  [Erratum-ibid.\ C {\bf 71}, 069902 (2005)].


\bibitem{Slac94}
L.~Andivahis {\it et al.},
%``Measurements of the electric and magnetic form-factors of the proton from
%Q**2 = 1.75-GeV/c**2 to 8.83-GeV/c**2,''
Phys.\ Rev.\ D {\bf 50}, 5491 (1994).
%%CITATION = PHRVA,D50,5491;%%


\bibitem{Chr04}
M.~E.~Christy {\it et al.}  [E94110 Coll.],
%``Measurements of electron proton elastic cross sections for 0.4-(GeV/c)**2 <
%Q**2 < 5.5-(GeV/c)**2,''
Phys.\ Rev.\ C {\bf 70}, 015206 (2004). 
%[arXiv:nucl-ex/0401030].
%%CITATION = NUCL-EX 0401030;%%


\bibitem{Qattan:2004ht}
 I.~A.~Qattan {\it et al.},
  %``Precision Rosenbluth measurement of the proton elastic form factors,''
  Phys.\ Rev.\ Lett.\  {\bf 94}, 142301 (2005). 
  %[arXiv:nucl-ex/0410010].
  %%CITATION = NUCL-EX 0410010;%%


\bibitem{Mar68}
  J.~Mar {\it et al.},
  %``A Comparison Of Electron - Proton And Positron - Proton Elastic Scattering
  %At Four Momentum Transfers Up To 5.0-Gev/C**2,''
  Phys.\ Rev.\ Lett.\  {\bf 21}, 482 (1968).
  %%CITATION = PRLTA,21,482;%%


\bibitem{Hartwig:1975px}
  S.~Hartwig {\it et al.},
  %``Elastic Scattering Of Electrons And Positrons From Protons,''
  Lett.\ Nuovo Cim.\  {\bf 12}, 30 (1975).
  %%CITATION = NCLTA,12,30;%%


\bibitem{MoTsai68}
L.~W.~Mo and Y.~S.~Tsai,
%``Radiative Corrections To Elastic And Inelastic E P And Mu P Scattering,''
Rev.\ Mod.\ Phys.\  {\bf 41}, 205 (1969).
%%CITATION = RMPHA,41,205;%%


\bibitem{oldyennie}
N. Meister and D. R. Yennie, Phys. Rev. {\bf 130}, 1210 (1963).


\bibitem{drell57}
S.~D.~Drell and M.~A.~Ruderman, 
Phys.\ Rev. {\bf 106}, 561 (1957). 


\bibitem{drell59}
S.~D.~Drell and S. Fubini, 
Phys.\ Rev. {\bf 113}, 741 (1959). 


\bibitem{werthamer61}
N.~R.~Werthamer and M.~A.~Ruderman, 
Phys.\ Rev. {\bf 123}, 1005 (1961). 


\bibitem{Greenhut:1970aq}
  G.~K.~Greenhut,
  %``Two-photon exchange in electron-proton scattering,''
  Phys.\ Rev.\  {\bf 184}, 1860 (1969).
  %%CITATION = PHRVA,184,1860;%%


\bibitem{GV03}
P.~A.~M.~Guichon and M.~Vanderhaeghen,
%``How to reconcile the Rosenbluth and the polarization transfer method in  the
%measurement of the proton form factors,''
Phys.\ Rev.\ Lett.\  {\bf 91}, 142303 (2003). 
%[arXiv:hep-ph/0306007].
%%CITATION = HEP-PH 0306007;%%


\bibitem{Arrington}
  J.~Arrington,
  %``Evidence for two photon exchange contributions in electron proton and
  %positron proton elastic scattering,''
  Phys.\ Rev.\ C {\bf 69}, 032201 (2004);
  %[arXiv:nucl-ex/0311019].
  %%CITATION = NUCL-EX 0311019;%%
  %``Extraction of two-photon contributions to the proton form factors,''
  Phys.\ Rev.\ C {\bf 71}, 015202 (2005). 
  %[arXiv:hep-ph/0408261].
  %%CITATION = HEP-PH 0408261;%%


\bibitem{Ji:1998pc}
  X.~D. Ji,
  %``Off-forward parton distributions,''
  J.\ Phys.\ G {\bf 24}, 1181 (1998).
  %[arXiv:hep-ph/9807358].
  %%CITATION = HEP-PH 9807358;%%


\bibitem{GPV01}
K.~Goeke, M.~V.~Polyakov and M.~Vanderhaeghen,
%``Hard exclusive reactions and the structure of hadrons,''
Prog.\ Part.\ Nucl.\ Phys.\  {\bf 47}, 401 (2001).
%[arXiv:hep-ph/0106012].
%%CITATION = HEP-PH 0106012;%%


\bibitem{Diehl:2003ny}
  M. Diehl,
  %``Generalized parton distributions,''
  Phys.\ Rept.\  {\bf 388}, 41 (2003).
  %[arXiv:hep-ph/0307382].
  %%CITATION = HEP-PH 0307382;%%


\bibitem{Ji:2004gf}
 X.~D. Ji,
  %``Generalized parton distributions,''
  Ann.\ Rev.\ Nucl.\ Part.\ Sci.\  {\bf 54}, 413 (2004).
  %%CITATION = ARNUA,54,413;%%


\bibitem{Belitsky:2005qn}
A.~V. Belitsky and A.~V.~Radyushkin,
  %``Unraveling hadron structure with generalized parton distributions,''
  Phys.\ Rept.\  {\bf 418}, 1 (2005).
  %[arXiv:hep-ph/0504030].
  %%CITATION = HEP-PH 0504030;%%



\bibitem{Eides:2000xc}
  M.~I.~Eides, H.~Grotch and V.~A.~Shelyuto,
  %``Theory of light hydrogenlike atoms,''
  Phys.\ Rept.\  {\bf 342}, 63 (2001). 
  %[arXiv:hep-ph/0002158].
  %%CITATION = HEP-PH 0002158;%%


\bibitem{Karshenboim:2005iy}
  S.~G.~Karshenboim,
  %``Precision physics of simple atoms: QED tests, nuclear structure and
  %fundamental constants,''
  Phys.\ Rept.\  {\bf 422}, 1 (2005). 
  %[arXiv:hep-ph/0509010].
  %%CITATION = HEP-PH 0509010;%%


\bibitem{Feinberg:1989ps}
  G.~Feinberg, J.~Sucher and C.~K.~Au,
  %``The Dispersion Theory Of Dispersion Forces,''
  Phys.\ Rept.\  {\bf 180}, 83 (1989).
  %%CITATION = PRPLC,180,83;%%


\bibitem{Barut60} 
A.O. Barut and C. Fronsdal,  Phys.\ Rev.\ {\bf 120}, 1871 (1960).


\bibitem{RKR71}
A.~De Rujula, J.~M.~Kaplan and E.~De Rafael,
%``Elastic Scattering Of Electrons From Polarized Protons And Inelastic Electron
%Scattering Experiments,''
Nucl.\ Phys.\ B {\bf 35}, 365 (1971).
%%CITATION = NUPHA,B35,365;%%



%%%%%%%%%%%%%%%%%%%%%%%%%%%
%
%      For 2gamma calculations 
%
%%%%%%%%%%%%%%%%%%%%%%%%%%%



\bibitem{BMT03}
P.~G.~Blunden, W.~Melnitchouk and J.~A.~Tjon,
%``Two-photon exchange and elastic electron proton scattering,''
Phys.\ Rev.\ Lett.\  {\bf 91}, 142304 (2003).
%[arXiv:nucl-th/0306076].
%%CITATION = NUCL-TH 0306076;%%


\bibitem{Blunden:2005ew}
  P.~G.~Blunden, W.~Melnitchouk and J.~A.~Tjon,
  %``Two-photon exchange in elastic electron nucleon scattering,''
  Phys.\ Rev.\ C {\bf 72}, 034612 (2005). 
  %[arXiv:nucl-th/0506039].
  %%CITATION = NUCL-TH 0506039;%%


\bibitem{Kondratyuk:2005kk}
  S.~Kondratyuk, P.~G.~Blunden, W.~Melnitchouk and J.~A.~Tjon,
  %``Delta resonance contribution to two-photon exchange in electron proton
  %scattering,''
  Phys.\ Rev.\ Lett.\  {\bf 95}, 172503 (2005). 
  %[arXiv:nucl-th/0506026].
  %%CITATION = NUCL-TH 0506026;%%


\bibitem{Kondratyuk:2007hc}
  S.~Kondratyuk and P.~G.~Blunden,
  %``Contribution of spin 1/2 and 3/2 resonances to two-photon exchange effects
  %in elastic electron proton scattering,''
  arXiv:nucl-th/0701003.
  %%CITATION = NUCL-TH 0701003;%%


\bibitem{Borisyuk:2006fh}
  D.~Borisyuk and A.~Kobushkin,
  %``Box diagram in the elastic electron-proton scattering,''
  Phys.\ Rev.\ C {\bf 74}, 065203 (2006); 
  %[arXiv:nucl-th/0606030].
  %%CITATION = NUCL-TH 0606030;%%
%``Two-photon exchange at low Q**2,''
  arXiv:nucl-th/0612104.
  %%CITATION = NUCL-TH 0612104;%%


\bibitem{YCC04}
Y.~C.~Chen, A.~Afanasev, S.~J.~Brodsky, C.~E.~Carlson and M.~Vanderhaeghen,
%``Partonic calculation of the two-photon exchange contribution to elastic
%electron proton scattering at large momentum transfer,''
Phys.\ Rev.\ Lett.\  {\bf 93}, 122301 (2004).
%[arXiv:hep-ph/0403058].
%%CITATION = HEP-PH 0403058;%%


\bibitem{ABCCV05}
 A.~V.~Afanasev, S.~J.~Brodsky, C.~E.~Carlson, Y.~C.~Chen and M.~Vanderhaeghen,
  %``The two-photon exchange contribution to elastic electron nucleon
  %scattering at large momentum transfer,''
  Phys.\ Rev.\ D {\bf 72}, 013008 (2005).
  %[arXiv:hep-ph/0502013].
  %%CITATION = HEP-PH 0502013;%%
 

\bibitem{oldkrass}
Allan S. Krass, Phys. Rev. {\bf 125}, 2172 (1962).


\bibitem{Ent:2001hm}
R.~Ent, B.~W.~Filippone, N.~C.~R.~Makins, R.~G.~Milner, T.~G.~O'Neill and D.~A.~Wasson,
%``Radiative corrections for (e,e-primep) reactions at GeV energies,''
Phys.\ Rev.\ C {\bf 64}, 054610 (2001).
%%CITATION = PHRVA,C64,054610;%%


\bibitem{Goldb57}
M.L. Goldberger, Y. Nambu and R. Oehme, Ann. of Phys. \textbf{2}, 226 (1957).


\bibitem{Rekalo:2003xa}
  M.~P.~Rekalo and E.~Tomasi-Gustafsson,
  %``Model independent properties of two-photon exchange in elastic electron
  %proton scattering,''
  Eur.\ Phys.\ J.\ A {\bf 22}, 331 (2004). 
  %[arXiv:nucl-th/0307066].
  %%CITATION = NUCL-TH 0307066;%%


\bibitem{Nie71}
P.~Van Nieuwenhuizen,
%``Muon-Electron Scattering Cross-Section To Order Alpha-To-The-Third,''
Nucl.\ Phys.\ B {\bf 28}, 429 (1971).
%%CITATION = NUPHA,B28,429;%%


\bibitem{grammer}
G.~J.~Grammer and D.~R.~Yennie,
%``Improved Treatment For The Infrared Divergence Problem In Quantum
%Electrodynamics,''
Phys.\ Rev.\ D {\bf 8}, 4332 (1973).
%%CITATION = PHRVA,D8,4332;%%


\bibitem{Brodsky:1968ea}
S.~J.~Brodsky and J.~R.~Primack,
%``The Electromagnetic Interactions Of Composite Systems,''
Annals Phys.\  {\bf 52}, 315 (1969).
%%CITATION = APNYA,52,315;%%
 

\bibitem{MT00}
L.~C.~Maximon and J.~A.~Tjon,
%``Radiative corrections to electron proton scattering,''
Phys.\ Rev.\ C {\bf 62}, 054320 (2000).
%[arXiv:nucl-th/0002058].
%%CITATION = NUCL-TH 0002058;%%


\bibitem{guidal} 
  M.~Guidal, M.~V.~Polyakov, A.~V.~Radyushkin and M.~Vanderhaeghen,
  %``Nucleon form factors from generalized parton distributions,''
  Phys.\ Rev.\ D {\bf 72}, 054013 (2005).
%  [arXiv:hep-ph/0410251].
  %%CITATION = HEP-PH 0410251;%%


\bibitem{diehl}
  M. Diehl, T.~Feldmann, R.~Jakob and P.~Kroll,
  %``Generalized parton distributions from nucleon form factor data,''
  Eur.\ Phys.\ J.\ C {\bf 39}, 1 (2005).
  %[arXiv:hep-ph/0408173].
  %%CITATION = HEP-PH 0408173;%%


\bibitem{Brash:2001qq}
  E.~J.~Brash, A.~Kozlov, S.~Li and G.~M.~Huber,
  %``New empirical fits to the proton electromagnetic form factors,''
  Phys.\ Rev.\ C {\bf 65}, 051001 (2002). 
  %[arXiv:hep-ex/0111038].
  %%CITATION = HEP-EX 0111038;%%


\bibitem{charles} 
JLab experiment E-04-019, spokespersons R. Gilman, L. Pentchev, C. Perdrisat, and R. Suleiman.


\bibitem{brooks}
Jefferson Lab experiment E-04-116; contact person, W.~Brooks.


\bibitem{Bystritskiy:2006ju}
  Yu.~M.~Bystritskiy, E.~A.~Kuraev and E.~Tomasi-Gustafsson,
  %``Application of the structure function method to polarized and unpolarized
  %electron proton scattering,''
  arXiv:hep-ph/0603132.
  %%CITATION = HEP-PH 0603132;%%


\bibitem{afanasev2005}
  The calculated result at 6 GeV$^2$ was presented by A. Afanasev at  the
  Workshop on Precision ElectroWeak Interactions held at William and  Mary,
  Williamsburg, VA, USA, 15-17 August 2005.  The related calculation  of hard bremsstrahlung
  effects in the polarized case was presented in
A.~Afanasev, I.~Akushevich and N.~Merenkov,
%``Model independent radiative corrections in processes of polarized   electron
%nucleon elastic scattering,''
Phys.\ Rev.\ D {\bf 64}, 113009 (2001).
%[arXiv:hep-ph/0102086].
%%CITATION = HEP-PH 0102086;%% 



\bibitem{Dong:2006wm}
  Y.~B.~Dong, C.~W.~Kao, S.~N.~Yang and Y.~C.~Chen,
  %``Observables in elastic electron deuteron scattering with two-photon
  %exchange,''
  Phys.\ Rev.\ C {\bf 74}, 064006 (2006). 
  %[arXiv:nucl-th/0610094].
  %%CITATION = NUCL-TH 0610094;%%


\bibitem{Rekalo:1999mt}
  M.~P.~Rekalo, E.~Tomasi-Gustafsson and D.~Prout,
  %``Search for evidence of two photon exchange in new experimental high
  %momentum transfer data on electron deuteron elastic scattering,''
  Phys.\ Rev.\ C {\bf 60}, 042202 (1999).
  %%CITATION = PHRVA,C60,042202;%%


\bibitem{Bordes:1986km}
  J.~Bordes, J.~A.~Penarrocha and J.~Bernabeu,
  %``TWO PHOTON EXCHANGE FORWARD AMPLITUDE IN ELASTIC LEPTON HADRON
  %SCATTERING,''
  Phys.\ Rev.\ D {\bf 35}, 3310 (1987).
  %%CITATION = PHRVA,D35,3310;%%


\bibitem{Penarrocha:1980fx}
  J.~A.~Penarrocha and J.~Bernabeu,
  %``Low Momentum Transfer Theorem In Lepton Hadron Scattering,''
  Annals Phys.\  {\bf 135}, 321 (1981).
  %%CITATION = APNYA,135,321;%%


\bibitem{Bernabeu:1980bi}
  J.~Bernabeu and J.~A.~Penarrocha,
  %``Two Photon Exchange In Muon - Nuclear Scattering,''
  Phys.\ Rev.\ D {\bf 22}, 1082 (1980).
  %%CITATION = PHRVA,D22,1082;%%


\bibitem{Arrington:2006hm}
  J.~Arrington and I.~Sick,
  %``Precise determination of low-Q nucleon electromagnetic form factors and
  %their impact on parity-violating e p elastic scattering,''
  arXiv:nucl-th/0612079.
  %%CITATION = NUCL-TH 0612079;%%


\bibitem{Afanasev:2005ex}
  A.~V.~Afanasev and C.~E.~Carlson,
  %``Two-photon-exchange correction to parity-violating elastic electron  proton
  %scattering,''
  Phys.\ Rev.\ Lett.\  {\bf 94}, 212301 (2005). 
  %[arXiv:hep-ph/0502128].
  %%CITATION = HEP-PH 0502128;%%


\bibitem{Blunden:2005jv}
  P.~G.~Blunden and I.~Sick,
  %``Proton radii and two-photon exchange,''
  Phys.\ Rev.\ C {\bf 72}, 057601 (2005). 
  %[arXiv:nucl-th/0508037].
  %%CITATION = NUCL-TH 0508037;%%


%%%%%%%%%%%%%%%%%%%%%%%%%%%
%
%      For hfs section
%
%%%%%%%%%%%%%%%%%%%%%%%%%%%


\bibitem{Karshenboim:1997zu}
S.~G.~Karshenboim,
%``What do we actually know about the proton radius?,''
Can.\ J.\ Phys.\  {\bf 77}, 241 (1999).
%[arXiv:hep-ph/9712347].
%%CITATION = HEP-PH 9712347;%%


\bibitem{Volotka:2004zu}
A.~V.~Volotka, V.~M.~Shabaev, G.~Plunien and G.~Soff,
%``Zemach and magnetic radius of the proton from the hyperfine splitting in
%hydrogen,''
Eur.\ Phys.\ J.\ D {\bf 33}, 23 (2005).
%[arXiv:physics/0405118].
%%CITATION = PHYS-ICS 0405118;%%

\bibitem{dupays}
A.~Dupays, A.~Beswick, B.~Lepetit, C.~Rizzo, and D.~Bakalov,
Phys.\ Rev.\ A {\bf 68}, 052503 (2003).


\bibitem{Eides:1995sq}
  M.~I.~Eides,
  %``Weak interaction contributions to hyperfine splitting and Lamb shift,''
  Phys.\ Rev.\ A {\bf 53}, 2953 (1996).
  %%CITATION = PHRVA,A53,2953;%%
  
 
\bibitem{Brodsky:2004ck}
  S.~J.~Brodsky, C.~E.~Carlson, J.~R.~Hiller and D.~S.~Hwang,
  %``Constraints on proton structure from precision atomic physics
  %measurements,''
  Phys.\ Rev.\ Lett.\  {\bf 94}, 022001 (2005);
  Phys.\ Rev.\ Lett.\ {\bf 94}, 169902 (E) (2005).
  %[arXiv:hep-ph/0408131].  
  %%CITATION = HEP-PH 0408131;%%


\bibitem{Friar:2005rs}
  J.~L.~Friar and I.~Sick,
  %``Comment on 'Constraints on proton structure from precision atomic  physics
  %measurements',''
  Phys.\ Rev.\ Lett.\  {\bf 95}, 049101 (2005);
  %[arXiv:nucl-th/0503020] 
  %%CITATION = NUCL-TH 0503020;%%
%\cite{Brodsky:2005gz}
%\bibitem{Brodsky:2005gz}
  S.~J.~Brodsky, C.~E.~Carlson, J.~R.~Hiller and D.~S.~Hwang,
  %``Brodsky et al. reply,''
  Phys.\ Rev.\ Lett.\  {\bf 95}, 049102 (2005).
  %%CITATION = PRLTA,95,049102;%%
  

\bibitem{Bodwin:1987mj}
G.~T.~Bodwin and D.~R.~Yennie,
%``Some Recoil Corrections To The Hydrogen Hyperfine Splitting,''
Phys.\ Rev.\ D {\bf 37}, 498 (1988).
%%CITATION = PHRVA,D37,498;%%


\bibitem{Martynenko:2004bt}
  A.~P.~Martynenko,
  %``2S hyperfine splitting of muonic hydrogen,''
  Phys.\ Rev.\ A {\bf 71}, 022506 (2005). 
  %[arXiv:hep-ph/0409107].
  %%CITATION = HEP-PH 0409107;%%

\bibitem{Zemach} 
A.~C.~Zemach, Phys.\ Rev.\ {\bf 104}, 1771 (1956).


\bibitem{Anthony:2000fn}
  P.~L.~Anthony {\it et al.}  [E155 Coll.],
  %``Measurements of the Q**2 dependence of the proton and neutron spin
  %structure functions g1(p) and g1(n),''
  Phys.\ Lett.\ B {\bf 493}, 19 (2000). 
  %[arXiv:hep-ph/0007248].
  %%CITATION = HEP-PH 0007248;%%
  

\bibitem{Anthony:2002hy}
  P.~L.~Anthony {\it et al.}  [E155 Coll.],
  %``Precision measurement of the proton and deuteron spin structure  functions
  %g2 and asymmetries A(2),''
  Phys.\ Lett.\ B {\bf 553}, 18 (2003). 
  %[arXiv:hep-ex/0204028].
  %%CITATION = HEP-EX 0204028;%%


\bibitem{Fatemi:2003yh}
  R.~Fatemi {\it et al.}  [CLAS Coll.],
  %``Measurement of the proton spin structure function g1(x,Q**2) for Q**2  from
  %0.15-GeV**2 to 1.6-GeV**2 with CLAS,''
  Phys.\ Rev.\ Lett.\  {\bf 91}, 222002 (2003). 
  %[arXiv:nucl-ex/0306019].
  %%CITATION = NUCL-EX 0306019;%%

\bibitem{models}
J.~Yun {\it et al.}, Phys.\ Rev.\ C {\bf 67} 055204, (2003); S. Kuhn, private communication.

\bibitem{deur}
A.~Deur, arXiv:nucl-ex/0507022.


\bibitem{Iddings} 
C.~K.~Iddings, Phys.\ Rev.\ {\bf 138}, B446 (1965).


\bibitem{Drell:1966kk}
S.~D.~Drell and J.~D.~Sullivan,
%``Polarizability Contribution To The Hydrogen Hyperfine Structure,''
Phys.\ Rev.\  {\bf 154}, 1477 (1967).
%%CITATION = PHRVA,154,1477;%%


\bibitem{DeRafael:mc}
E.~De Rafael,
%``The Hydrogen Hyperfine Structure And Inelastic Electron Proton Scattering
%Experiments,''
Phys.\ Lett.\ B {\bf 37}, 201 (1971).
%%CITATION = PHLTA,B37,201;%%


\bibitem{Gnaedig:qt}
P.~Gn\"adig and J.~Kuti,
%``Proton Structure And Hyperfine Splitting In The Hydrogen Atom,''
Phys.\ Lett.\ B {\bf 42}, 241 (1972).
%%CITATION = PHLTA,B42,241;%%


\bibitem{Faustov:yp}
R.~N.~Faustov and A.~P.~Martynenko,
%``Proton Polarizability Contribution To Hydrogen Hyperfine Splitting,''
Eur.\ Phys.\ J.\ C {\bf 24}, 281 (2002);
%%CITATION = EPHJA,C24,281;%%
R.~N.~Faustov and A.~P.~Martynenko,
%``Proton polarizability contribution to the hydrogen hyperfine
%splitting,''
Phys.\ Atom.\ Nucl.\  {\bf 65}, 265 (2002)
[Yad.\ Fiz.\  {\bf 65}, 291 (2002)].
%[arXiv:hep-ph/0007044].
%%CITATION = HEP-PH 0007044;%%


\bibitem{Nazaryan:2005zc}
  V.~Nazaryan, C.~E.~Carlson and K.~A.~Griffioen,
  % ``New experimental constraints on polarizability corrections to hydrogen
  %hyperfine structure,''
  Phys.\ Rev.\ Lett.\  {\bf 96}, 163001 (2006). 
  %[arXiv:hep-ph/0512108].
  %%CITATION = HEP-PH 0512108;%%


\bibitem{Gerasimov:1965et}
  S.~B.~Gerasimov,
  %``A Sum Rule For Magnetic Moments And The Damping Of The Nucleon Magnetic
  %Moment In Nuclei,''
  Sov.\ J.\ Nucl.\ Phys.\  {\bf 2}, 430 (1966)
  [Yad.\ Fiz.\  {\bf 2}, 598 (1966)].
  %%CITATION = SJNCA,2,430;%%


\bibitem{Drell:1966jv}
  S.~D.~Drell and A.~C.~Hearn,
  %``Exact Sum Rule For Nucleon Magnetic Moments,''
  Phys.\ Rev.\ Lett.\  {\bf 16}, 908 (1966).
  %%CITATION = PRLTA,16,908;%%
  

\bibitem{Simula:2001iy}
  S.~Simula, M.~Osipenko, G.~Ricco and M.~Taiuti,
  %``Leading and higher twists in the proton polarized structure function  g1(p)
  %at large Bjorken x,''
  Phys.\ Rev.\ D {\bf 65}, 034017 (2002). 
  %[arXiv:hep-ph/0107036].  
Silvano Simula provided us with an updated version of the code, 
including error estimates.
%%CITATION = HEP-PH 0107036;%%


\bibitem{Faustov:2006ve}
  R.~N.~Faustov, I.~V.~Gorbacheva and A.~P.~Martynenko,
  %``Proton polarizability effect in the hyperfine splitting of the hydrogen
  %atom,''
  arXiv:hep-ph/0610332.
  %%CITATION = HEP-PH 0610332;%%
  

\bibitem{Karshenboim:1996ew}
S.~G.~Karshenboim,
%``Nuclear structure-dependent radiative corrections to the hydrogen  hyperfine
%splitting,''
Phys.\ Lett.\  {\bf 225A}, 97 (1997).
%[arXiv:hep-ph/9608484].
%%CITATION = HEP-PH 9608484;%%


\bibitem{Friar:2003zg}
J.~L.~Friar and I.~Sick,
%``Zemach Moments for Hydrogen and Deuterium,''
Phys.\ Lett.\ B {\bf 579}, 285 (2004).
%[arXiv:nucl-th/0310043].
%%CITATION = NUCL-TH 0310043;%%


\bibitem{Mohr:2000ie}
P.~J.~Mohr and B.~N.~Taylor,
%``CODATA recommended values of the fundamental physical constants: 1998,''
Rev.\ Mod.\ Phys.\  {\bf 72}, 351 (2000);
%%CITATION = RMPHA,72,351;%%
and Rev.\ Mod.\ Phys.\ {\bf 77}, 1 (2005) [2002 CODATA]; {\it Cf.}, the electron scattering value of  I.~Sick,
  %``On the rms-radius of the proton,''
  Phys.\ Lett.\ B {\bf 576}, 62 (2003)
  %[arXiv:nucl-ex/0310008] 
  %%CITATION = NUCL-EX 0310008;%%
  or the spread from Sick's value to that of
J.~J.~Kelly,
  %``Simple parametrization of nucleon form factors,''
  Phys.\ Rev.\ C {\bf 70}, 068202 (2004).
  %%CITATION = PHRVA,C70,068202;%%


\bibitem{Drechsel:2002ar}
  D.~Drechsel, B.~Pasquini and M.~Vanderhaeghen,
  %``Dispersion relations in real and virtual Compton scattering,''
  Phys.\ Rept.\  {\bf 378}, 99 (2003).
  %[arXiv:hep-ph/0212124].
  %%CITATION = HEP-PH 0212124;%%


\bibitem{Pedroni:2006ta}
  P.~Pedroni  [GDH and A2 Coll.],
  %``Recent results on the GDH sum rule on neutron and proton,''
  AIP Conf.\ Proc.\  {\bf 814}, 374 (2006);
  %%CITATION = APCPC,814,374;%%
 H.~Dutz {\it et al.}  [GDH Coll.],
  % ``Measurement of helicity-dependent photoabsorption cross sections on the
  %neutron from 815-MeV to 1825-MeV,''
  Phys.\ Rev.\ Lett.\  {\bf 94}, 162001 (2005).
  %%CITATION = PRLTA,94,162001;%%


\bibitem{Dicus:2000cd}
  D.~A.~Dicus and R.~Vega,
  %``The Drell-Hearn sum rule at order alpha**3,''
  Phys.\ Lett.\ B {\bf 501}, 44 (2001). 
  %[arXiv:hep-ph/0011212];
  %%CITATION = HEP-PH 0011212;%%


\bibitem{Altarelli:1972nc}
  G.~Altarelli, N.~Cabibbo and L.~Maiani,
  %``The Drell-Hearn sum rule and the lepton magnetic moment in the Weinberg
  %model of weak and electromagnetic interactions,''
  Phys.\ Lett.\ B {\bf 40}, 415 (1972).
  %%CITATION = PHLTA,B40,415;%%
  
\bibitem{abarb67} 
  H. D. I. Abarbanel and S. Nussinov, Phys.\ Rev.\ {\bf 158}, 1462 (1967).

\bibitem{Damashek:1969xj}
  M.~Damashek and F.~J.~Gilman,
  %``Forward Compton Scattering,''
  Phys.\ Rev.\ D {\bf 1}, 1319 (1970);
  %%CITATION = PHRVA,D1,1319;%%
  C.~A.~Dominguez, C.~Ferro Fontan and R.~Suaya,
  %``Forward Compton scattering sum rules and j=0 singularities,''
  Phys.\ Lett.\ B {\bf 31}, 365 (1970).
  %%CITATION = PHLTA,B31,365;%%


\bibitem{Brodsky:1973hm}
  S.~J.~Brodsky, F.~E.~Close and J.~F.~Gunion,
  %``A Gauge - Invariant Scaling Model Of Current Interactions With Regge
  %Behavior And Finite Fixed Pole Sum Rules,''
  Phys.\ Rev.\ D {\bf 8}, 3678 (1973).
  %%CITATION = PHRVA,D8,3678;%%


\bibitem{Antognini:2005fe}
  A.~Antognini {\it et al.},
  %``The 2S Lamb shift in muonic hydrogen and the proton rms charge radius,''
  AIP Conf.\ Proc.\  {\bf 796}, 253 (2005).
  %%CITATION = APCPC,796,253;%%


\bibitem{Faustov:2001pn}
  R.~N.~Faustov, E.~V.~Cherednikova and A.~P.~Martynenko,
  % ``Proton polarizability contribution to the hyperfine splitting in muonic
  %hydrogen,''
  Nucl.\ Phys.\ A {\bf 703}, 365 (2002). 
  %[arXiv:hep-ph/0108044].
  %%CITATION = HEP-PH 0108044;%%


%%%%%%%%%%%%%%%%%%%%%%%%%%%
%
%      For section on beam and target normal spin asymmetries
%
%%%%%%%%%%%%%%%%%%%%%%%%%%%



\bibitem{RKR73}
A. De Rujula, J.M. Kaplan, and E. de Rafael, 
Nucl. Phys. {\bf B 53}, 545 (1973).


\bibitem{Wells01}
S.P. Wells {\it et al.} (SAMPLE Coll.), 
Phys. Rev. C {\bf 63}, 064001 (2001). 


\bibitem{Maas03}
 F.~E.~Maas {\it et al.} (A4 Coll.),
  %``Measurement of the transverse beam spin asymmetry in elastic electron
  %proton scattering and the inelastic contribution to the imaginary part  of
  %the two-photon exchange amplitude,''
  Phys.\ Rev.\ Lett.\  {\bf 94}, 082001 (2005). 
  %[arXiv:nucl-ex/0410013].
  %%CITATION = NUCL-EX 0410013;%%


\bibitem{Kaufmann}
L. Kaufmann (on behalf of Happex Coll.), proceedings PAVI06. 


\bibitem{Capozza}
L. Capozza (on behalf of A4 Coll.), proceedings PAVI06.


\bibitem{E158}
K. Kumar (on behalf of E-158 Coll.), proceedings PAVI06. 


\bibitem{G0}
S.K. Phillips, P. King (G0 Coll.), 
  DNP meeting Oct 25-28 2006, Nashville TN, 
  paper HC.00013


\bibitem{AAM02}
A. Afanasev, I. Akusevich, and N.P. Merenkov, 
hep-ph/0208260.


\bibitem{GGV04}
M.~Gorchtein, P.~A.~M.~Guichon and M.~Vanderhaeghen,
  %``Beam normal spin asymmetry in elastic lepton nucleon scattering,''
  Nucl.\ Phys.\ A {\bf 741}, 234 (2004).
  %[arXiv:hep-ph/0404206].
  %%CITATION = HEP-PH 0404206;%%


\bibitem{Tar75}
R. Tarrach, Nuovo Cimento A {\bf 28}, 409 (1975).


\bibitem{Diaconescu:2004aa}
  L.~Diaconescu and M.~J.~Ramsey-Musolf,
 Phys.\ Rev.\ C {\bf 70}, 054003 (2004)


\bibitem{Pasquini:2004pv}
B.~Pasquini and M.~Vanderhaeghen,
  %``Resonance estimates for single spin asymmetries in elastic electron
  %nucleon scattering,''
  Phys.\ Rev.\ C {\bf 70}, 045206 (2004).
  %[arXiv:hep-ph/0405303].
  %%CITATION = HEP-PH 0405303;%%


\bibitem{Dre99}
D. Drechsel, O. Hanstein, S. Kamalov, and L. Tiator,
Nucl. Phys. {\bf A645}, 145 (1999).


\bibitem{Afanasev:2004pu}
  A.~V.~Afanasev and N.~P.~Merenkov,
  Phys.\ Lett.\ B {\bf 599}, 48 (2004); 
for corrected formulas and corrected numerical results see also 
  [arXiv:hep-ph/0407167].


\bibitem{Gorchtein:2005za}
  M.~Gorchtein,
  Phys.\ Rev.\ C {\bf 73}, 035213 (2006);
  Phys.\ Rev.\ C {\bf 73}, 055201 (2006).
 

\bibitem{Gorchtein:2006mq}
M.~Gorchtein,
  %``Dispersive contributions to e+ p/e- p cross section ratio in forward
  %regime,''
  Phys.\ Lett.\ B {\bf 644}, 322 (2007).
  %[arXiv:hep-ph/0610378].
  %%CITATION = HEP-PH 0610378;%%


\bibitem{AnTodd} 
JLab experiment E-05-015, spokespersons T. Averett, J.P. Chen, X. Jiang.


%%%%%%%%%%%%%%%%%%%%%%%%%%%%%%%%%%%%%%%%%%%%%%%%%%%%%%%%%%%%%%%%%
%
% Conclusion, outlook
%
%%%%%%%%%%%%%%%%%%%%%%%%%%%%%%%%%%%%%%%%%%%%%%%%%%%%%%%%%%%%%%%%%


\bibitem{Tvaskis:2005ex}
  V.~Tvaskis {\it et al.},
  %``Experimental constraints on non-linearities induced by two-photon  effects
  %in elastic and inelastic Rosenbluth separations,''
  Phys.\ Rev.\ C {\bf 73}, 025206 (2006)
  %[arXiv:nucl-ex/0511021].
  %%CITATION = NUCL-EX 0511021;%%


\bibitem{arrington2}
J. Arrington, V.F. Dimitriev,R.J. Holt, D.M. Nikolenko, L.A.
Rachek, Yu.V. Shestakov, V.N. Stibunov, D.K. Toporkov, H. de Vries; 
Proposal for experiment
at VEPP-3, nucl-ex/0408020.

\bibitem{Pascalutsa:2005es}
  V.~Pascalutsa, C.~E.~Carlson and M.~Vanderhaeghen,
  %``Two-photon exchange effects in the electro-excitation of the Delta
  %resonance,''
  Phys.\ Rev.\ Lett.\  {\bf 96}, 012301 (2006).
  %[arXiv:hep-ph/0509055].
  %%CITATION = HEP-PH 0509055;%%


\bibitem{Bartels:1973pf}
  J.~Bartels,
  %``Contributions Of Multi - Photon Exchange In E P Deep Inelastic
  %Scattering,''
  Nucl.\ Phys.\ B {\bf 82}, 172 (1974).
  %%CITATION = NUPHA,B82,172;%%


\bibitem{Kingsley:1972fd}
  R.~L.~Kingsley,
  %``Two-photon exchange effects in deep inelastic electroproduction,''
  Nucl.\ Phys.\ B {\bf 46}, 615 (1972); 
  %%CITATION = NUPHA,B46,615;%%
%\bibitem{Fishbane:1974cd}
  P.~M.~Fishbane and R.~L.~Kingsley,
  %``Difference between e+ and e- deep-inelastic scattering,''
  Phys.\ Rev.\ D {\bf 8}, 3074 (1973).
  %%CITATION = PHRVA,D8,3074;%%


\bibitem{Bodwin:1975ty}
  G.~T.~Bodwin and C.~D.~Stockham,
  %``Comment On 'Difference Between E+ And E- Deep Inelastic Scattering',''
  Phys.\ Rev.\ D {\bf 11}, 3324 (1975).
  %%CITATION = PHRVA,D11,3324;%%



\bibitem{Marciano:1982mm}
  W.~J.~Marciano and A.~Sirlin,
  %``Radiative Corrections To Atomic Parity Violation,''
  Phys.\ Rev.\ D {\bf 27}, 552 (1983); 
  %%CITATION = PHRVA,D27,552;%%
%\bibitem{Marciano:1983ss}
%  W.~J.~Marciano and A.~Sirlin,
  %``On Some General Properties Of The O (Alpha) Corrections To Parity Violation
  %In Atoms,''
  Phys.\ Rev.\ D {\bf 29}, 75 (1984)
  [Erratum-ibid.\ D {\bf 31}, 213 (1985)].
  %%CITATION = PHRVA,D29,75;%%


\bibitem{Bohm:1986na}
  M.~Bohm and H.~Spiesberger,
  %``Radiative Corrections To Neutral Current Deep Inelastic Lepton Nucleon
  %Scattering At Hera Energies,''
  Nucl.\ Phys.\ B {\bf 294}, 1081 (1987).
  %%CITATION = NUPHA,B294,1081;%%





\end{thebibliography}
\end{document}